%% file: main_AJ.tex
\documentclass[twocolumn, times]{aastex631}

\usepackage{xspace}
\usepackage{url}
\usepackage{color}
\usepackage{amsmath}

\newcommand\TY{2021~TY$_{14}$\xspace}
\newcommand\UW{2021~UW$_{1}$\xspace}
\newcommand\GQ{2022~GQ$_{1}$\xspace}
\newcommand\radTY{15\xspace}

\newcommand\radGQ{10\xspace}
\newcommand\radstara{20\xspace}
\newcommand\radstarb{10\xspace}
\newcommand\Nmc{3000\xspace}
\newcommand\rotPofTY{$15.281\pm0.002$~s\xspace}
\newcommand\rotPofUW{$21.099\pm0.003$~s\xspace}
\newcommand\rotPofGQ{$8.779\pm0.013$~s\xspace}

\newcommand\abminTY{1.51\xspace}

\newcommand\abminUW{1.14\xspace}

\newcommand\abminGQ{1.24\xspace}

\newcommand\dCgrTY{0.098\xspace}
\newcommand\dCriTY{0.040\xspace}
\newcommand\dCgrUW{0.019\xspace}
\newcommand\dCriUW{0.015\xspace}
\newcommand\dCgrGQ{0.217\xspace}
\newcommand\dCriGQ{0.280\xspace}
\newcommand\dCgrupperTY{0.057\xspace}
\newcommand\dCriupperTY{0.044\xspace}
\newcommand\dCgrupperUW{0.058\xspace}
\newcommand\dCriupperUW{0.041\xspace}
\newcommand\dCgrupperGQ{0.457\xspace}
\newcommand\dCriupperGQ{0.309\xspace}
\newcommand\fspotmaxTY{0.22\xspace}
\newcommand\fspotmaxUW{0.21\xspace}

\received{March 27, 2026}
\revised{May 18, 2026}
\accepted{June 2, 2026}

\submitjournal{AJ}

\shorttitle{Tricolor Video Observations of Tiny NEAs}
\shortauthors{Beniyama et al.}

\begin{document}
\title{
   Simultaneous Tricolor Video Observations of Three Tiny Near-Earth Asteroids with Sub-Minute Rotation Periods
}


\correspondingauthor{Jin Beniyama}
\email{jbeniyama@oca.eu}

\author[0000-0003-4863-5577]{Jin Beniyama}
\affiliation{Université Côte d'Azur, 
    Observatoire de la Côte d'Azur, CNRS, Laboratoire Lagrange, Bd de l'Observatoire, 
    CS 34229, 06304 Nice Cedex 4, France
}
\affiliation{
    Department of Earth and Planetary Science, The University of Tokyo, 7-3-1 Hongo, Bunkyo, Tokyo 113-0033, Japan
}
\author[0000-0001-5797-6010]{Ryou Ohsawa}
\affiliation{
    National Astronomical Observatory of Japan, 2-21-1 Osawa, Mitaka, Tokyo 181-8588, Japan
}
\affiliation{
    Astronomical Science Program, Graduate Institute for Advanced Studies, SOKENDAI, 2-21-1 Osawa, Mitaka, Tokyo 181-8588, Japan
}
\author[0000-0002-8792-2205]{Shigeyuki Sako}
\affiliation{
    Institute of Astronomy, Graduate School of Science, The University of Tokyo, 2-21-1 Osawa, Mitaka, Tokyo 181-0015, Japan
}
\affiliation{
    UTokyo Organization for Planetary Space Science, The University of Tokyo, 7-3-1 Hongo, Bunkyo-ku, Tokyo 113-0033, Japan
}
\affiliation{
    Collaborative Research Organization for Space Science and Technology, The University of Tokyo, 7-3-1 Hongo, Bunkyo-ku,
Tokyo 113-0033, Japan
}
\affiliation{
    Research Center for the Early Universe, Graduate School of Science, The University of Tokyo, 7-3-1 Hongo, Bunkyo-ku, 
Tokyo, 113-0033, Japan
}
\affiliation{
    Next-generation Neutrino Science and Multi-messenger Astronomy Organization, The University of Tokyo, 5-1-5 Kashiwanoha, 
Kashiwa, Chiba 277-8582, Japan
}
\author{Satoshi Takita}
\affiliation{
    Institute of Astronomy, Graduate School of Science, The University of Tokyo, 2-21-1 Osawa, Mitaka, Tokyo 181-0015, Japan
}
\author[0000-0003-1726-6158]{Tomohiko Sekiguchi}
\affiliation{
    Hokkaido University of Education, 9 Hokumon, Asahikawa, 070-8621, Japan
}
\author[0000-0002-7363-187X]{Daisuke Kuroda}
\affiliation{
    Japan Spaceguard Association,
    Bisei Spaceguard Center 1716-3 Okura, Bisei, Ibara, Okayama 714-1411, Japan
}
\author[0000-0002-6480-3799]{Keisuke Isogai}
\affiliation{
    Okayama Observatory, Kyoto University, 3037-5 Honjo, Kamogatacho, Asakuchi, Okayama 719-0232, Japan
}
\affiliation{
    Department of Multi-Disciplinary Sciences, Graduate School of Arts and Sciences, The University of Tokyo, 3-8-1 Komaba, Meguro, Tokyo 153-8902, Japan
}

\begin{abstract}
Studying the physical properties of near-Earth asteroids (NEAs) is crucial for understanding their dynamical histories and origins, 
and assessing impact hazards to Earth.
Tiny NEAs with diameters smaller than 100~m are intrinsically faint and 
are typically observable only during close approaches,
resulting in few well-characterized objects.
Furthermore, because these objects are often fast-moving and fast-rotating, sequential multiband photometry 
is prone to systematic offsets in derived colors.
To mitigate this effect, we
performed simultaneous $g$-, $r$-, and $i$-band photometry of three tiny NEAs using the TriColor CMOS Camera and Spectrograph (TriCCS) on the 3.8~m Seimei Telescope.
We used high-cadence video observations with exposure times of 1~s and 5~s to investigate lightcurve variations on timescales of seconds.
All three NEAs are confirmed as fast rotators with rotation periods shorter than 60~s:
$15.281\pm0.002$~s for 2021~TY$_{14}$,
$21.099\pm0.003$~s for 2021~UW$_{1}$,
and
$8.779\pm0.013$~s for 2022~GQ$_{1}$.
The derived colors indicate that 2021~TY$_{14}$ belongs to the X-complex, while 2021~UW$_{1}$ and 2022~GQ$_{1}$ belong to the S-complex.
Their positions in the diameter–rotation period diagram show that all three objects belong to the small,
fast-rotating NEA population,
with 2022~GQ$_{1}$ being the smallest and fastest-rotating among them with spectroscopic measurements.
Analysis of the color time series suggests that the surfaces of observed NEAs are largely homogeneous, although 2021~TY$_{14}$ exhibits statistically significant $g-r$ color heterogeneity with a projected spot fraction of approximately 50\%. For 2021~UW$_{1}$, minor localized variations of up to $\sim20$\% in composition cannot be ruled out.
Our study demonstrates that
high-cadence, multicolor photometry with small to medium telescopes is effective for determining 
rotation periods and broadly classifying the spectral properties of tiny NEAs.
\end{abstract}

\section{Introduction} \label{sec:intro}
Studying near-Earth asteroids (NEAs) is crucial for understanding their dynamical histories and origins, as well as for planetary defense efforts to mitigate potential asteroid impacts on Earth.
One of the fundamental physical quantities of asteroids is their rotation period.
In the gravity-dominated regime, where self-gravity exceeds material strength, 
asteroids with rotation periods shorter than about 2~hr, the so-called cohesionless spin barrier, 
are thought to undergo structural failure or disruption \citep{Pravec2000b}.
Therefore, the rotation period provides important constraints on an asteroid’s internal structure.

Many authors have investigated 
fast-rotating asteroids that rotate faster than the cohesionless spin barrier in various asteroid populations,
including main-belt asteroids 
\citep{Chang2014b, Chang2016, Chang2017, Chang2019, Yeh2020, Chang2022b, Strauss2024, Novakovic2025}, Jovian Trojans \citep{Chang2021, Kiss2025}, and Hildas \citep{Chang2022a, Takacs2025}.
Recently, \citet{Greenstreet2026} derived reliable rotation periods of 75 main-belt asteroids and one NEA observed in the first LSST Camera commissioning images from the Vera C. Rubin Observatory.
The spin barriers, or equivalently the shortest rotation periods, differ among these population and likely reflect differences in composition.
Importantly, \citet{Carbognani2017} investigated the spectral types of fast-rotating asteroids with rotation periods close to the cohesionless spin barrier.
They showed that the observationally 
determined
critical rotation periods of S- and C-type asteroids in the 4--20~km size range differ by a factor of about 1.20.
This provides evidence for the existence of the cohesionless spin barrier, which can be explained by differences in bulk density between spectral types.

In recent years, an increasing number of rotation periods shorter 
than the cohesionless spin barrier have been reported for small NEAs, 
as well as for some MBAs, observed either through dedicated campaigns or serendipitously.
The Mission Accessible Near-Earth Objects Survey (MANOS) has carried out dedicated lightcurve observations, spectroscopy, and spectrophotometry of small NEAs \citep{Thirouin2016, Thirouin2018, Devogele2019, Moskovitz2026}.
The survey assembled lightcurves for 228 small NEAs using 1--4~m class telescopes \citep{Thirouin2016, Thirouin2018}.
The  sample includes four sub-minute rotators: 2014~RC with a rotation period of 15.8~s, 2015~SV$_6$ with 18~s, 2016~MA with 18.4~s, and 2017~QG$_{18}$ with 11.9~s.
Another statistical study of the physical properties of NEAs is a spectroscopic survey conducted with the NASA Infrared Telescope Facility \citep[IRTF,][]{Sanchez2024}.
That study combined spectroscopic data for 84 small NEAs obtained between 2017 and 2021.
Photometric data for 59 NEAs, obtained with a guide camera on the IRTF, were also analyzed.
Subsecond photometry, or video observations, of tiny NEAs with diameters below 100~m were conducted using the 1.05~m telescope at Kiso Observatory in Japan \citep{Beniyama2022}.
Rotation periods were derived for 32 NEAs, including 13 sub-minute rotators.
A dedicated survey of 249 NEAs using the Two-meter Twin Telescope system (TTT3, 2.0~m; TTT1 and TTT2, 0.8~m) and the 1~m Transient Survey Telescope (TST) identified 16 sub-minute rotators \citep{Alarcon2026}.

A target-of-opportunity streak photometry was conducted with 
the Canada--France--Hawaii Telescope (CFHT) on Mauna Kea, Hawaii \citep{Bolin2024}.
This study reported three sub-minute rotators from long-exposure photometry: 
2016~GE$_1$ with a rotation period of 31~s, 
2016~CG$_{18}$ with 55~s, 
and 2016~EV$_{84}$ with 52~s.
Two additional sub-minute rotators were identified using aperture photometry on asteroid trails \citep{Devogele2024b},
with possible rotation periods of 9.16~s and 18.33~s reported for 2023~CX$_1$ ($D\sim1$~m).
The fastest rotation period ever reported, $2.5888\pm0.0002$~s, was found for 2024~BX$_1$ \citep[$D\sim1$~m,][]{Devogele2024b}.
The Great Shefford Observatory (MPC code J95) also regularly reports fast-rotating asteroids including sub-minute rotators 
\citep{Birtwhistle2009, Birtwhistle2011a, Birtwhistle2011b, Birtwhistle2018a, Birtwhistle2018b, Birtwhistle2021a, Birtwhistle2021b, Birtwhistle2021c, Birtwhistle2021d, Birtwhistle2021e, Birtwhistle2022a, Birtwhistle2022b, Birtwhistle2022c, Birtwhistle2023a, Birtwhistle2023b, Birtwhistle2023c, Birtwhistle2023d, Birtwhistle2024a, Birtwhistle2024b, Birtwhistle2024c, Birtwhistle2024d, Birtwhistle2025a, Birtwhistle2025b, Birtwhistle2025c}.

Spin barriers, if present, can be identified in diameter–rotation period diagrams even for objects rotating faster than the cohesionless spin barrier.
To achieve a comprehensive physical understanding of these barriers, 
it is essential to characterize the physical properties of the objects, 
such as their taxonomic classification and surface composition.
Although many lightcurves of fast-rotating asteroids have been obtained, relatively few spectroscopic or spectrophotometric observations have been reported.
This might primarily be due to their faintness, the limited observing windows during close approaches, and the restricted availability of medium- to large-aperture telescopes.
Furthermore, when spectrophotometric measurements are obtained through sequential observations with multiple filters, 
the data are susceptible to rotational lightcurve variations during the sequence. 
These variations, combined with the necessity of correcting for different reference stars in each filter’s field of view, require rigorous calibration and can introduce significant systematic uncertainties. 
\citet{Polishook2012} performed both spectroscopic and lightcurve observations of two tiny NEAs, 2012~KP$_{24}$ ($D\sim20$~m) and 2012~KT$_{42}$ ($D\sim6$~m), 
as part of a rapid-response program at NASA's IRTF.
They derived rotation periods of approximately 2.5~min for 2012~KP$_{24}$ and 3.6~min for 2012~KT$_{42}$.
Two fast-rotators were characterized within the framework of the Visible Near-Earth Objects Survey \citep[ViNOS;][]{Licandro2023}.
That study reported rotation periods of about 13~min for 2021~NY$_1$ ($D\sim100$~m) and about 3~min for 2022~AB ($D\sim65$~m assuming a geometric albedo of 0.15).
More recently, the sub-10-meter-class NEA 2022~OB$_5$ was observed with the 10.4-m Gran Telescopio Canarias (GTC) and TTT3, and its rotation period was derived to be $1.542\pm0.001$~min \citep{Alarcon2026b}. 
Using HiPERCAM on the GTC, they obtained high-cadence simultaneous five-band observations in the $u$-, $g$-, $r$-, $i$-, and $z$-bands, and found that the spectrum is consistent with the X-complex.
The spectral-type distribution of fast-rotating asteroids remains only partially understood, and that of sub-minute rotators is even less well constrained.

We aim to demonstrate that high-cadence, simultaneous tricolor photometry can determine both rotation periods and visible colors of tiny fast-ratoting NEAs.
In this paper, we report the results of visible simultaneous multicolor photometry of three tiny NEAs.
The paper is organized as follows.
Section~2 describes our photometric observations and data reduction.
The observational results, along with constraints on the physical properties of the three NEAs and their implications, are presented in Section~3.

\section{Observations and data reduction}
\subsection{Tricolor video observations}
We conducted simultaneous multicolor photometric observations of three NEAs: 
\TY\ on October~15, 2021, \UW\ on October~29, 2021, and \GQ\ on April~7, 2022.
The observations of \TY\ and \GQ\ were carried out as part of a dual-epoch program focused on tiny NEAs (PI: Jin Beniyama), which is embedded within the Optical and Infrared Synergetic Telescopes for Education and Research (OISTER) collaboration.
The OISTER network operates small ground-based telescopes in Japan and South Africa under an inter-university framework.
The observations of \UW\ were obtained opportunistically during a gap in the observing schedule of a single-epoch program targeting the NEA (3200)~Phaethon \citep[PI: Tomohiko Sekiguchi;][]{Beniyama2023a}.

The observing circumstances are summarized in Table~\ref{tab:obs}.
All observations were performed with the TriColor CMOS Camera and Spectrograph (TriCCS) mounted on the 3.8~m Seimei Telescope \citep{Kurita2020}, located at the Kyoto University Okayama Observatory (longitude 133.5967$^\circ$~E, latitude 34.5769$^\circ$~N, altitude 355~m).
We simultaneously acquired images in three bands 
whose filter responses are similar to those of the Pan-STARRS $g$, $r$, and $i$ filters \citep{Chambers2016}.
The instrument provides a field of view of $12.6\arcmin\times7.5\arcmin$ with a pixel scale of 0.350~arcsec~pixel$^{-1}$.

\begin{figure*}
\centering
\begin{minipage}[b]{0.32\hsize}
    \centering
    \includegraphics[width=\hsize]{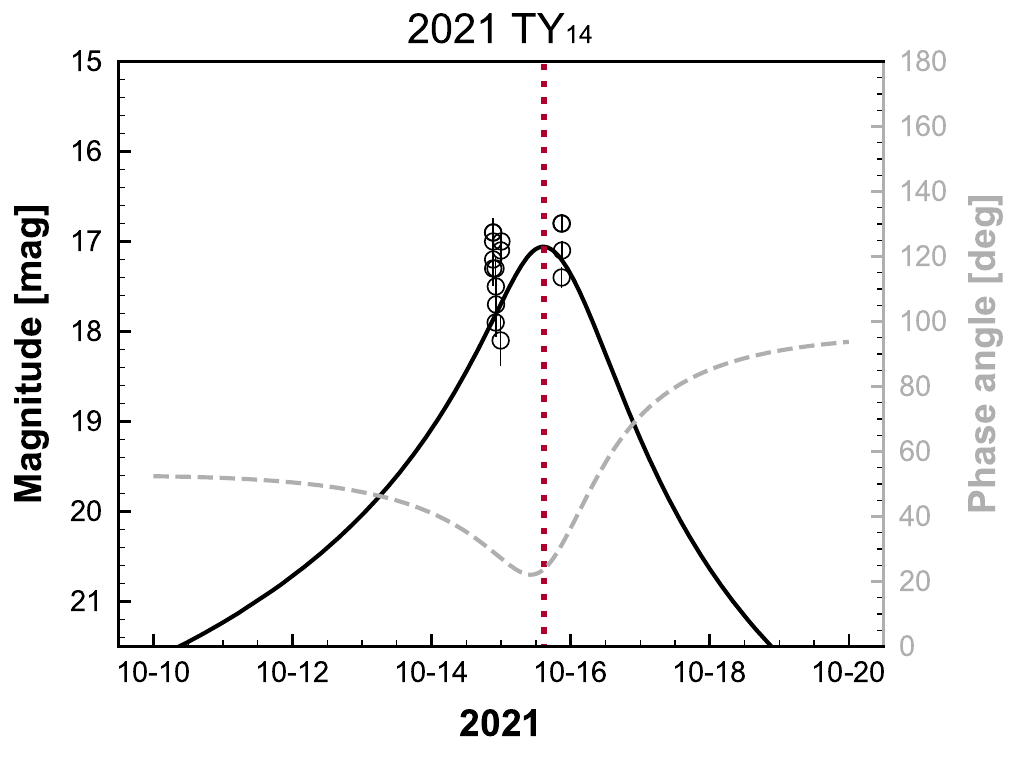}
\end{minipage}
\hfill
\begin{minipage}[b]{0.32\hsize}
    \centering
    \includegraphics[width=\hsize]{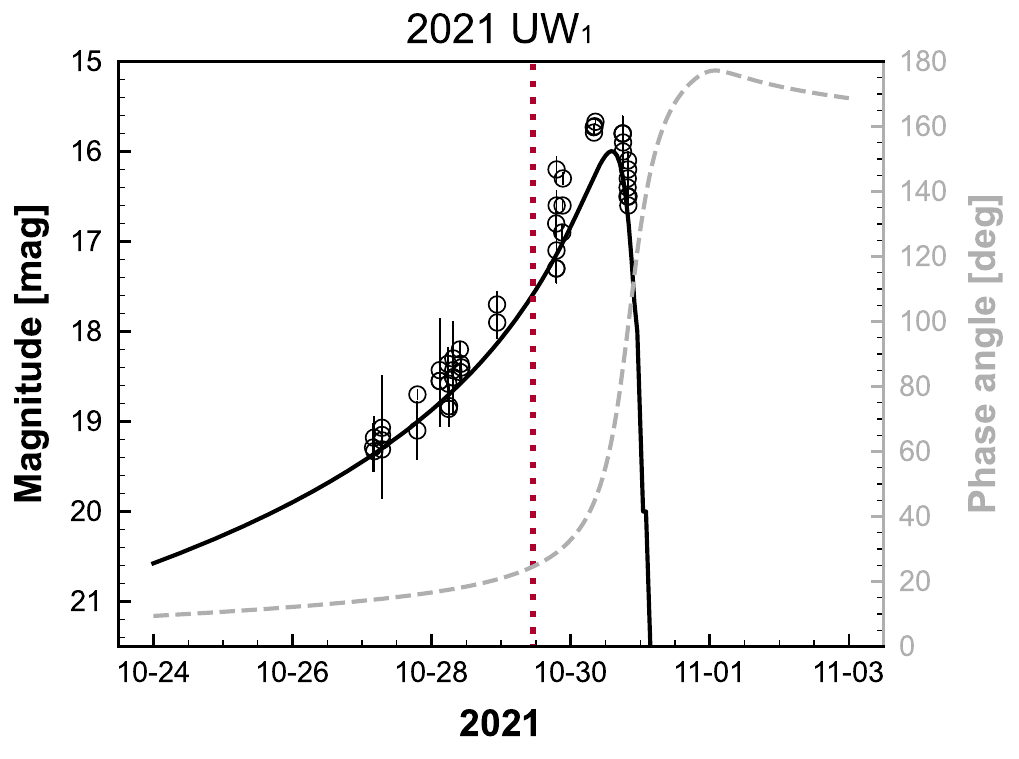}
\end{minipage}
\hfill
\begin{minipage}[b]{0.32\hsize}
    \centering
    \includegraphics[width=\hsize]{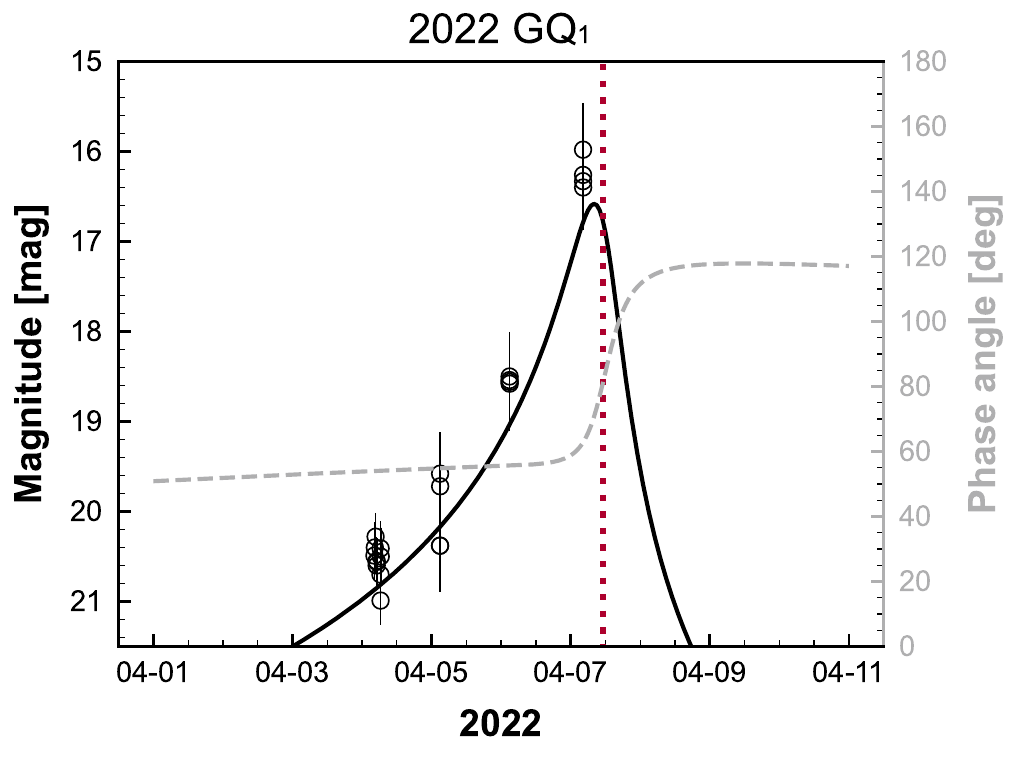}
\end{minipage}
\caption{
Ephemerides of \TY, \UW, and \GQ.
Apparent brightnesses and phase angles are shown by solid and dashed lines, respectively.
The times of our observations are indicated by vertical dotted lines.
Photometric data from the Minor Planet Center (MPC) database (\url{https://www.minorplanetcenter.net/db_search}, accessed on October 31, 2025) with
reported magnitude uncertainties are shown as circles with the error bars.
For illustrative purposes, 
all available photometric data are plotted regardless of the calibrated band.
}
\label{fig:ephem}
\end{figure*}

The apparent brightnesses and solar phase angles of the observed NEAs are shown in Fig.~\ref{fig:ephem}.
Observations were carried out near the epochs of peak apparent brightness for the asteroids.
All observations were carried out with nonsidereal tracking at the apparent motion rates of the targets.
The exposure times were set to 
1~s for both \TY\ and \GQ, and 5~s for \UW.
Rather than using single long exposures, we obtained a sequence of short-exposure images.
This strategy minimizes the elongation of background reference stars caused by nonsidereal tracking and enables us to resolve rapid lightcurve variations associated with the rotation of the NEAs.

\subsection{Data reduction}
We applied standard image reduction procedures, including bias subtraction, dark subtraction, and flat-field correction, to all frames.
Astrometric solutions were derived using reference sources from the Gaia Data Release~2 catalog and the \texttt{astrometry.net} software package \citep{Lang2010}.
A small fraction of the data was discarded owing to poor image quality, unstable observing conditions, 
or cases in which the target overlapped with nearby sources, preventing reliable photometry.

The colors and magnitudes of the NEAs were derived following the same procedures as those used in \citet{Beniyama2023a, Beniyama2023b, Beniyama2023c, Beniyama2024, Beniyama2025b, Beniyama2025c}, except for the photometry of \GQ, which required a different approach (see below).
Cosmic rays were identified and removed using the Python package \texttt{astroscrappy} \citep{McCully2018}, which implements the L.A.Cosmic algorithm developed by \citet{vanDokkum2001}.
Circular aperture photometry of the NEAs was performed using the SExtractor-based Python package \texttt{sep}.
The aperture radii were set to \radTY~pix for \TY and \UW, corresponding to approximately 1.5 times the full width at half maximum (FWHM) of the point-spread functions (PSFs) of nearby reference stars.
Photometry of the reference stars was also carried out using \texttt{sep}.
The same aperture radii as those used for the NEAs were applied to the reference stars.

Photometric calibration was performed using the Pan-STARRS Data Release~2 catalog \citep{Chambers2016}.
Reference stars were excluded from the analysis if they met any of the following criteria:
catalog uncertainties in the $g$, $r$, or $i$ bands larger than 0.05~mag;
$(g-r)_{\mathrm{PS}} > 1.1$;
$(g-r)_{\mathrm{PS}} < 0.0$;
$(r-i)_{\mathrm{PS}} > 0.8$;
or $(r-i)_{\mathrm{PS}} < 0.0$,
where $(g-r)_{\mathrm{PS}}$ and $(r-i)_{\mathrm{PS}}$ denote colors in the Pan-STARRS photometric system.
In addition, photometric measurements obtained within 100~pixels of the image edges or affected by contamination from nearby sources within the aperture were rejected on a frame-by-frame basis.
Extended sources, candidate quasars, and variable stars were removed using the \texttt{objinfoflag} and \texttt{objfilterflag} parameters in the Pan-STARRS catalog.
Typically, an average of nine reference stars were available for each frame for both \TY and \UW.
To mitigate systematic errors in color and magnitude estimates, frames containing fewer than three stars were excluded from the analysis. 
The observed scatter in the magnitude zero points remained around 0.01 mag, corresponding to a conservative estimate of the systematic bias under the assumption that the photometric uncertainties of the reference stars are negligible.
Due to the high apparent velocity of the target, the field of view shifted several times during the observations, resulting in the use of different sets of reference stars for photometric calibration. 
However, we consider this effect to have a negligible impact on our results because each field contained a sufficient number of reference stars to ensure a stable photometric zero point.
A similar approach was employed by \citet{Beniyama2023a, Beniyama2023c}, where consistent colors were successfully derived from observations spanning different fields and multiple nights using the same instruments. 

Due to the sparsely populated field of \GQ, 
we adopted a different analysis strategy
The signal-to-noise ratio (S/N) of \GQ\ in individual 1~s frames is typically a few to less than ten in the $g$ band.
While this S/N is sufficient to detect lightcurve variations with amplitudes exceeding this level, 
no reference stars with adequate S/N were available in the field.
Therefore, we derived a relative instrumental lightcurve under the assumption that sky conditions remained stable over 
the observation duration of approximately 80~s, enabling us to examine rotational brightness variations.
We tested several aperture radii and adopted \radGQ~pix as the nominal value; the resulting lightcurve properties are insensitive to this choice.

To determine colors for \GQ, image stacking was required to increase the S/N of the reference stars.
We stacked 20 successive frames with individual exposure times of 1~s, resulting in frames with effective exposure times of 20~s.
The typical
duration time of pixel reset 
is approximately 0.4~ms, which is negligible compared to the total exposure time.
Two types of stacked frames were produced: 
nonsidereally stacked frames, which minimize the elongation of \GQ by using the World Coordinate System (WCS) solutions derived from surrounding field sources, 
and sidereally stacked frames, which minimize the elongation of reference star images.
Photometry of \GQ was obtained from the nonsidereally stacked frames, while photometry of the reference stars was measured from the sidereally stacked frames.
We adopted an aperture radius of \radGQ~pix for \GQ.
Because the images of the reference stars are elongated even in the 1~s exposures due to the large apparent motion of \GQ and the use of nonsidereal tracking, 
we used an elliptical aperture for reference stars, with semi-major and semi-minor axes of \radstara~pix and \radstarb~pix, respectively.
Two reference stars were used to account for color terms in the calibration.
In total, the first three frames were used in the analysis.
The last frame was excluded because the two reference stars were located outside the field of view.

\input{tab_phot}

\section{Results and discussion} \label{sec:result}
\subsection{Rotation periods and axial ratios} \label{subsec:res_lc}
\begin{figure*}
\centering
\begin{minipage}[b]{0.32\hsize}
    \centering
    \includegraphics[width=\hsize]{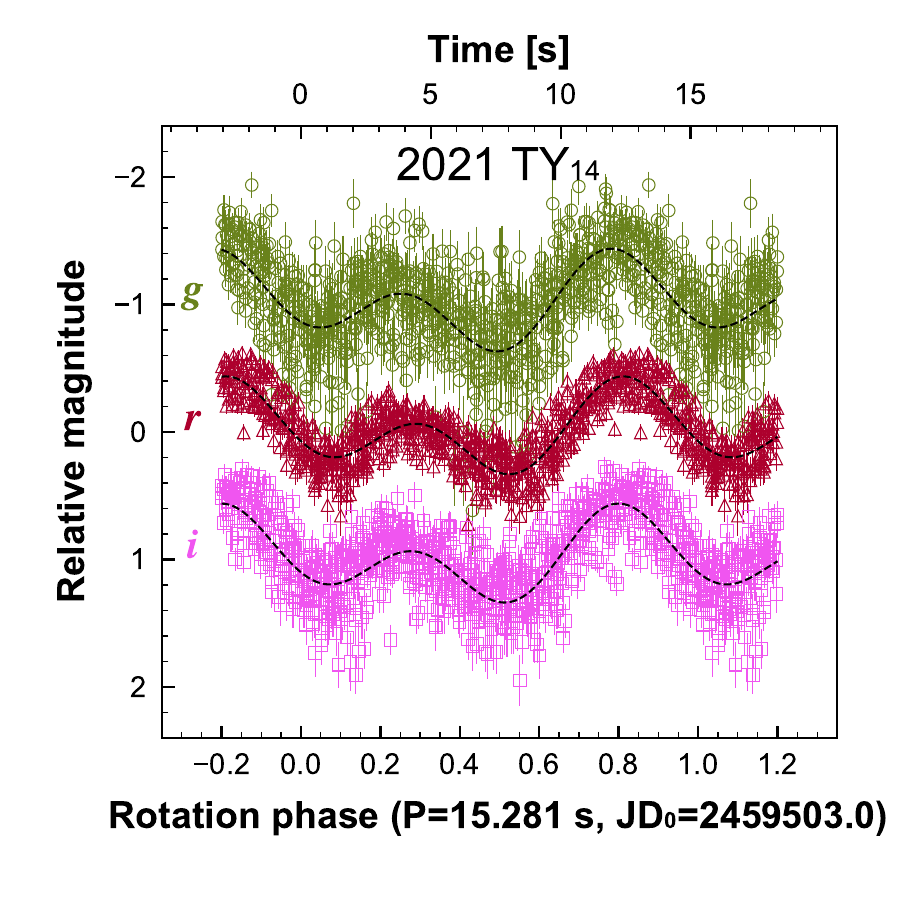}
\end{minipage}
\hfill
\begin{minipage}[b]{0.32\hsize}
    \centering
    \includegraphics[width=\hsize]{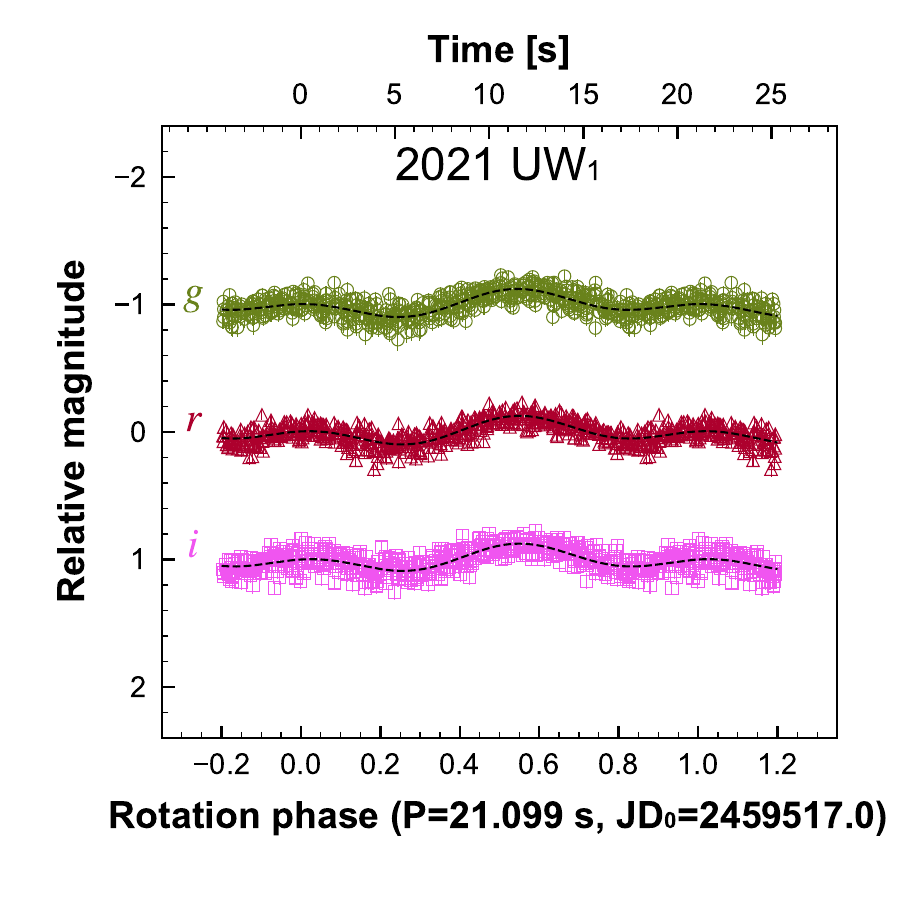}
\end{minipage}
\hfill
\begin{minipage}[b]{0.32\hsize}
    \centering
    \includegraphics[width=\hsize]{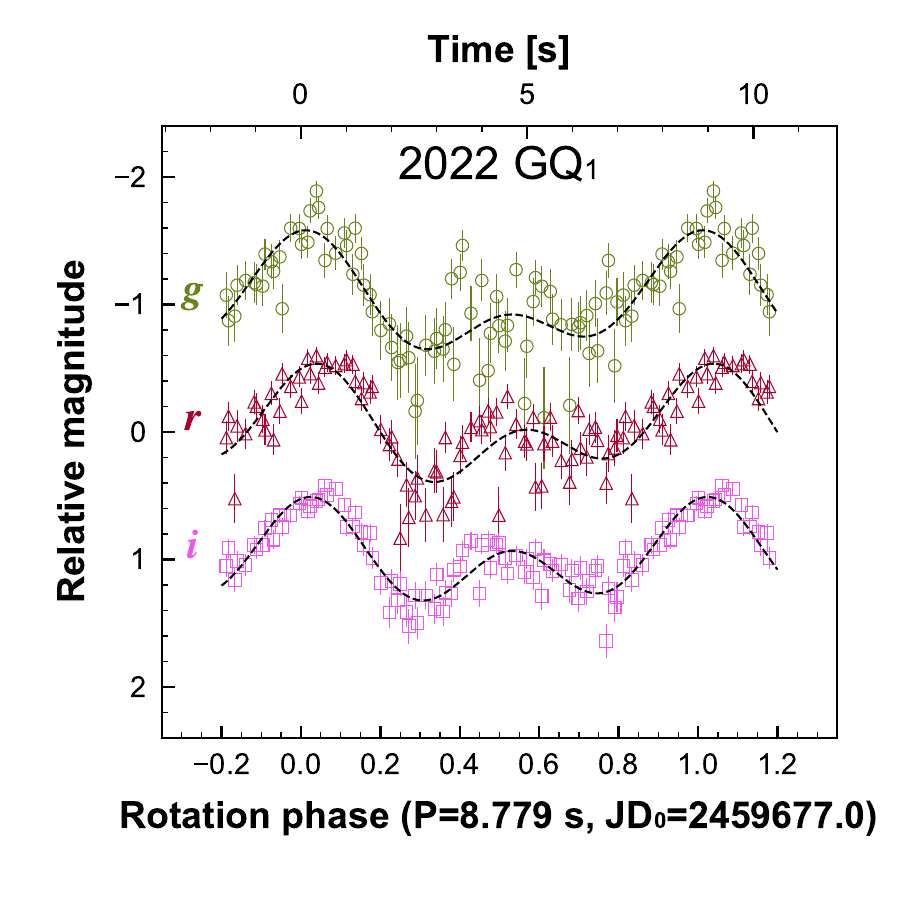}
\end{minipage}
\caption{
Phased lightcurves of \TY, \UW, and \GQ.
From top to bottom, the $g$, $r$, and $i$ bands are plotted using circles, triangles, and squares, respectively.
Phase zero is set to 
JD 2459503.0 (2021 October 15 12:00:00 UT) for \TY,
JD 2459504.5 (2021 October 29 12:00:00 UT) for \UW, 
and
JD 2459505.2 (2022 April 07 12:00:00 UT) for \GQ.
The $g$ and $i$ bands lightcurves are vertically offset by $-1$ and $+1$~mag, respectively.
Bars indicate the 1$\sigma$ photometric uncertainties.
Fitting model curves are shown by dashed lines.
}
\label{fig:plc}
\end{figure*}

The lightcurves of the NEAs are shown in Figs. \ref{fig:lc_full_TY}--\ref{fig:lc_full_GQ}.
These lightcurves exhibit clear variations.
We performed a periodic analysis using the lightcurves with the highest signal-to-noise ratio: the $r$-band lightcurves for \TY\ and \UW, and the $i$-band lightcurve for \GQ, applying the Lomb--Scargle technique \citep{Lomb1976, Scargle1982, VanderPlas2018}.
The details and results of periodic analyses are shown in Figs. \ref{fig:LS_TY}--\ref{fig:LS_MC_GQ}.
The derived rotation periods and lightcurve amplitudes are summarized in Table \ref{tab:res}.
Assuming the double-peak lightcurves,
we found the rotation periods of 
\rotPofTY for \TY,  \rotPofUW for \UW, and \rotPofGQ for \GQ.
The phased lightcurves are shown in Fig. \ref{fig:plc}.
Three asteroids are all sub-minute rotators with rotation periods less than 60~s.

\citet{Beniyama2022} performed video observations of \TY 
with an exposure time of 0.5~s
using the CMOS camera Tomo-e Gozen on October 15, 2021,
approximately three hours before our observations.
The rotation period was reported to be $15.292\pm0.002$~s,
and lightcurve amplitude was estimated to be $0.61\pm0.02$~mag.
Our results are broadly consistent with them. 
\citet{Birtwhistle2022a} reported observations of \UW on October 30, 2021 when \UW was as bright as 16~mag in $V$ band and 
its solar phase angle was as large as 90~deg (see Fig. \ref{fig:ephem}).
They obtained 43~minutes of lightcurves with 0.9~s exposure.
The rotation period with the strongest peak of the period spectrum was estimated to be 
$0.0058620\pm0.0000004$~hr, or $21.1032\pm0.0014$~s.
Their lightcurve amplitude was estimated to be as large as 1.2~mag.
Our rotation period is consistent with the solution in \citet{Birtwhistle2022a}.
Our smaller lightcurve amplitude is interpreted as a results of 
long exposure compared to the rotation period, 5~s, 
as well as the smaller solar phase angle of about 24~deg.
These consistencies indicate that our measurements are reliable.

The axial ratio of asteroids provides insight into their internal structure 
(e.g., monolithic bodies or rubble-piles) 
and possible formation mechanisms, since different formation processes 
(e.g., collisions or rotational fission) may produce distinct shape distributions.
By comparing the inferred axial ratios of our targets with previously reported axial ratio-diameter trends, 
we can examine whether these tiny sub-minute rotators show any unusual structural characteristics or instead follow the general population trend.
We assumed the asteroid is a triaxial ellipsoid with
axial lengths of $a$, $b$, and $c$ ($a \geq b \geq c$) and the aspect angle, angle between rotation axis and the asteroids--observer direction, of 90~deg. 
A lower limit on the axial ratio $a/b$ is given by:
\begin{equation}
(a/b)_\mathrm{min} = 10^{0.4 \Delta m(\alpha)/(1+m\alpha)}\,,
\end{equation}
where $\Delta m(\alpha)$ is the lightcurve amplitude at a phase angle of $\alpha$ 
and $m$ is a ratio of a slope of amplitude-phase relationship to $\Delta m(0)$, 
which is known to depend on the taxonomic type of the asteroid \citep{Bowell1989}.
When we assume $m$ of 0.030, a typical value of S-type asteroids \citep{Zappala1990}, which corresponds to the maximum slope \citep{Gutierrez2006}, 
we obtain conservative lower limits for the axial ratios:
$a/b >$ \abminTY for \TY,
$a/b >$ \abminUW for \UW, 
and
$a/b >$ \abminGQ for \GQ.
As discussed above, 
the amplitude for \UW\ was likely underestimated due to the smearing effect.
Thus, its axial ratio should be regarded as a strict lower limit.
These axial ratios of tiny sub-minute rotators are consistent with the trend reported in previous studies investigating the correlation between diameter, rotation period, and axial ratio \citep{Hatch2015, Thirouin2016, Beniyama2022}.

\input{tab_res}

\subsection{Surface colors}
\begin{figure}
\centering
\includegraphics[width=1.0\hsize]{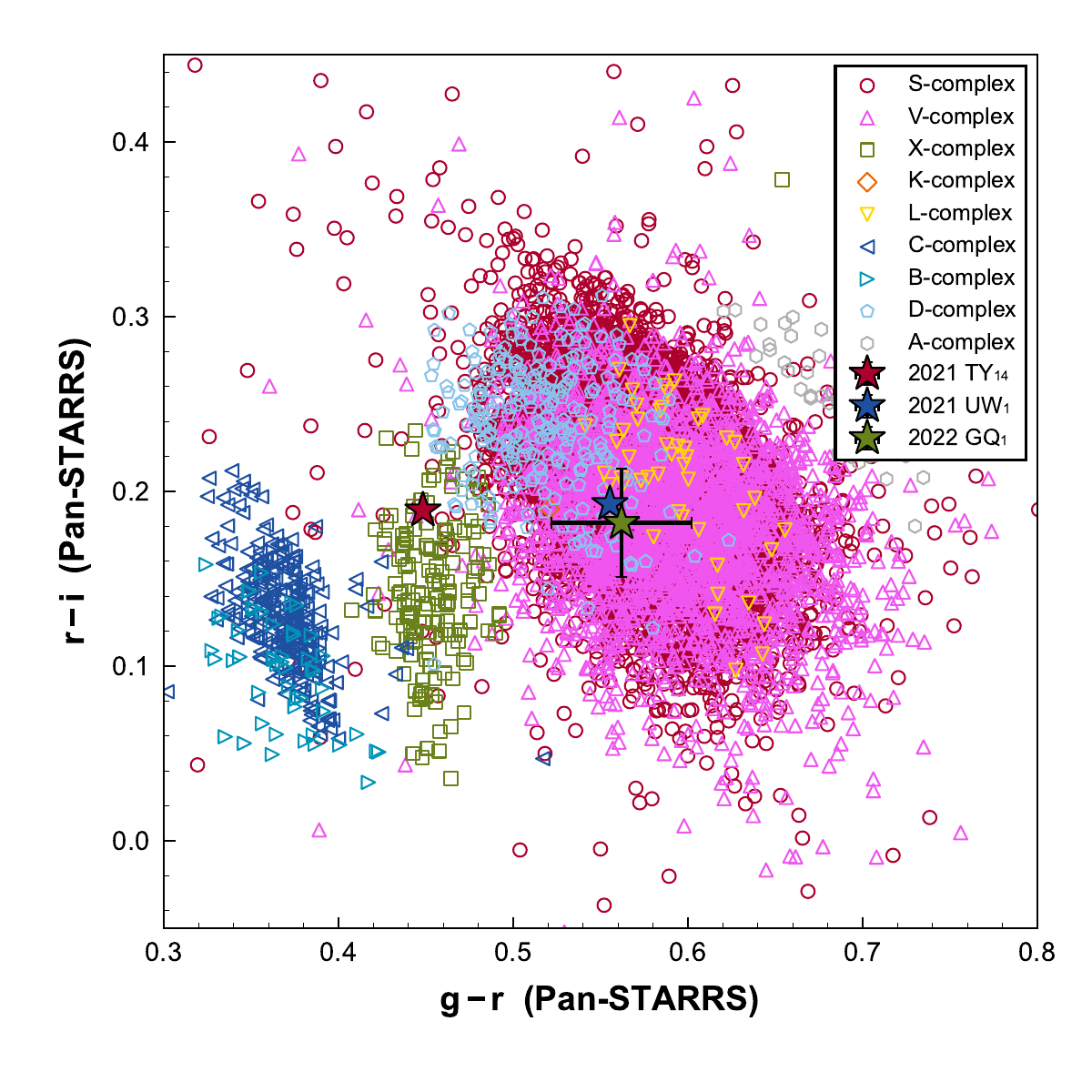}
\caption{
Color-color diagram of $g$-$r$ vs. $r$-$i$.
The weighted mean colors of the NEAs, derived from individual nightly measurements, are plotted as stars.
The error bars represent the uncertainties of the weighted means.
Asteroids from \citet{Sergeyev2021} are also plotted: 
S-complex (circles), 
V-complex (triangles),
X-complex (squares),
K-complex (diamonds),
L-complex (inverse triangles),
C-complex (left-pointing triangles),
B-complex (right-pointing triangles),
D-complex (pentagons),
and 
A-complex (hexagons). 
}
\label{fig:cc}
\end{figure}

The derived colors of the NEAs are summarized in Table~\ref{tab:res}.
The colors were calculated as the weighted mean of individual measurements, with uncertainties expressed as the standard error of the weighted mean.
Note that the total uncertainty may be dominated by systematic uncertainties in the photometric calibration (typically $\sim$0.01 mag), even when the statistical uncertainties are as small as 0.001 mag.
For comparison, the derived colors are plotted in Fig.~\ref{fig:cc} together with those of asteroids drawn from a recent color catalog \citep{Sergeyev2021}.
In that catalog, each asteroid is assigned probabilities of belonging to different taxonomic complexes.
We selected asteroids with probabilities of 80\% or higher for a given complex, excluding the U class, which denotes objects of unknown taxonomy.
We computed the $g-r$ and $r-i$ colors in the SDSS system and subsequently transformed them into the Pan-STARRS system using the conversion equations provided by \citet{Tonry2012}.
By a visual inspection of the color--color diagrams, 
\TY\ is consistent with the region occupied by X-complex asteroids, 
whereas \UW\ and \GQ\ overlap with the region of S-complex asteroids.

Color time series can be used to investigate their surface homogeneity \citep[e.g.,][]{Degewij1979}.
Various ranges of color differences have been reported in the literature:
0.06--0.11 mag for asteroids in the Sloan Digital Sky Survey Moving Object Catalog \citep{Szabo2004},
$\sim0.04$~mag for a TNO (136108)~2003 EL$_{61}$ \citep[also known as Haumea,][]{Lacerda2008},
and $\sim0.02$~mag for an NEA (3200)~Phaethon \citep{Beniyama2023a}.
Simultaneous photometry is a powerful technique for investigating surface homogeneity, as it reduces the effects of atmospheric variations on the measurements \citep[e.g.,][]{Beniyama2023a}.

\begin{figure}
\centering
\includegraphics[width=1.0\hsize]{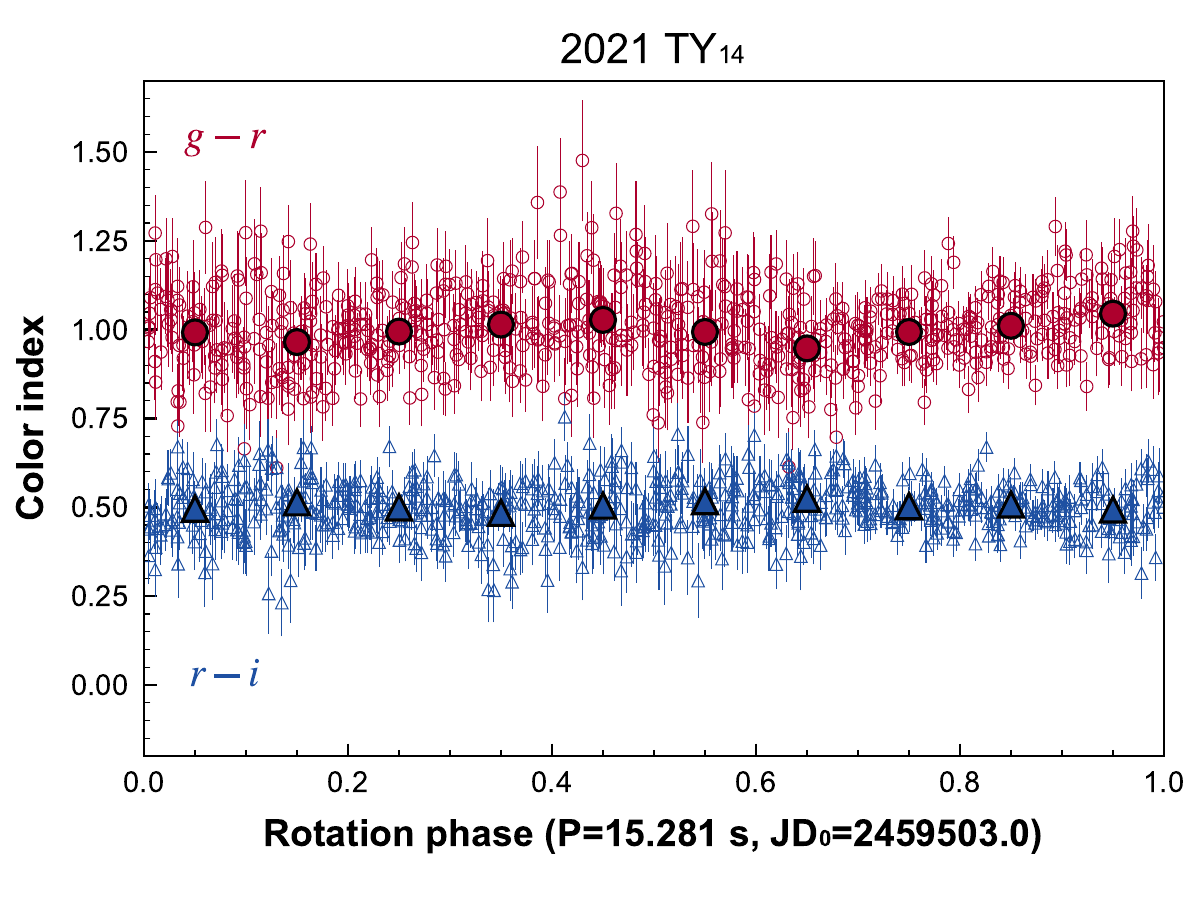}
\caption{
Phased color time series of \TY. 
The $g-r$ and $r-i$ color time series are presented in the top and bottom plots, indicated by circles and triangles, respectively.
Weighted average $g-r$ and $r-i$ colors 
per 0.1 phase are shown by filled circles and filled triangles, respectively.
The $g-r$ and $r-i$ color time series are normalized such that their median values are unity. 
The $r-i$ color time series is
vertically offset by 0.5~mag for clarity.
Phase zero is set to 
JD 2459503.0 (2021 October 15 12:00:00 UT). 
Bars indicate the 1$\sigma$ uncertainties.
}
\label{fig:col_TY}
\end{figure}

\begin{figure}
\centering
\includegraphics[width=1.0\hsize]{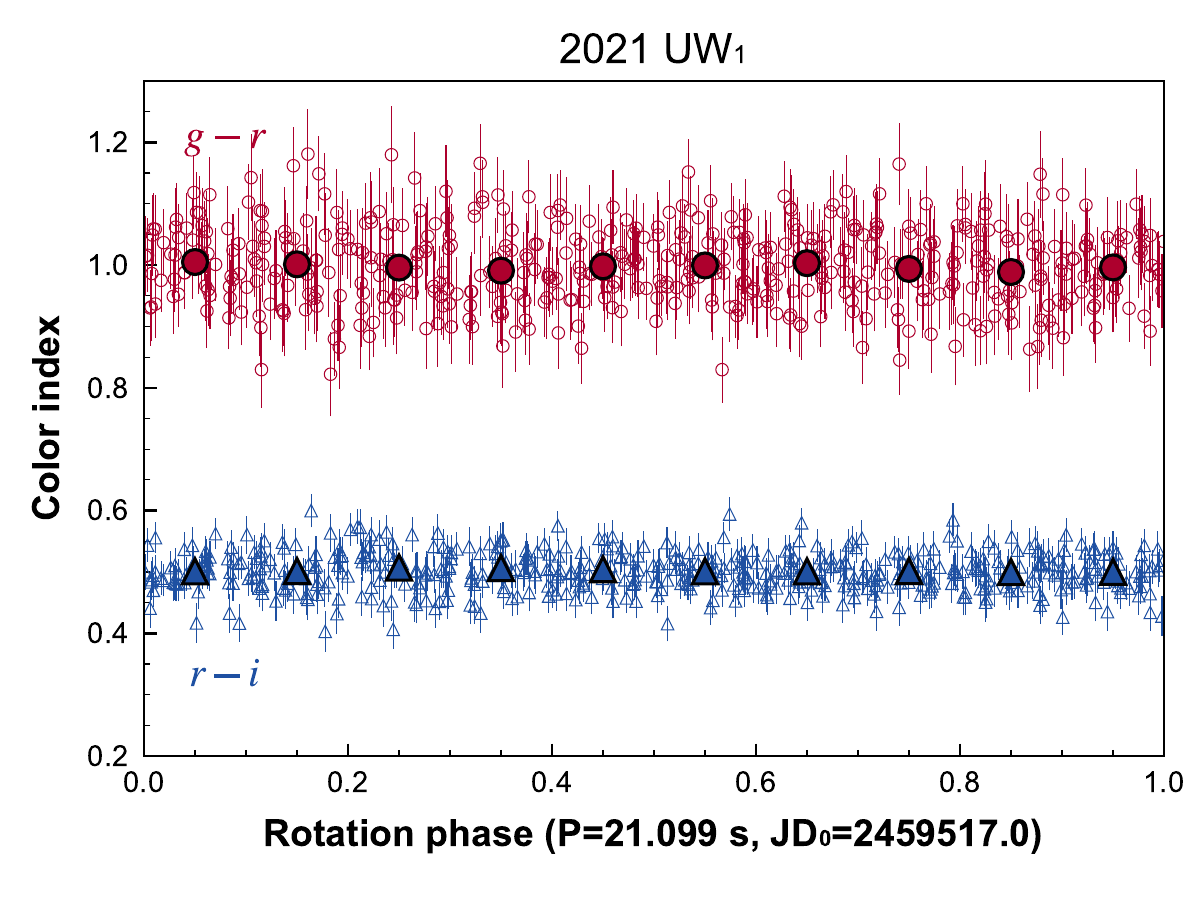}
\caption{
Same as Fig. \ref{fig:col_TY}, but for \UW.
Phase zero is set to  
JD 2459504.5 (2021 October 29 12:00:00 UT).
}
\label{fig:col_UW}
\end{figure}

\begin{figure}
\centering
\includegraphics[width=1.0\hsize]{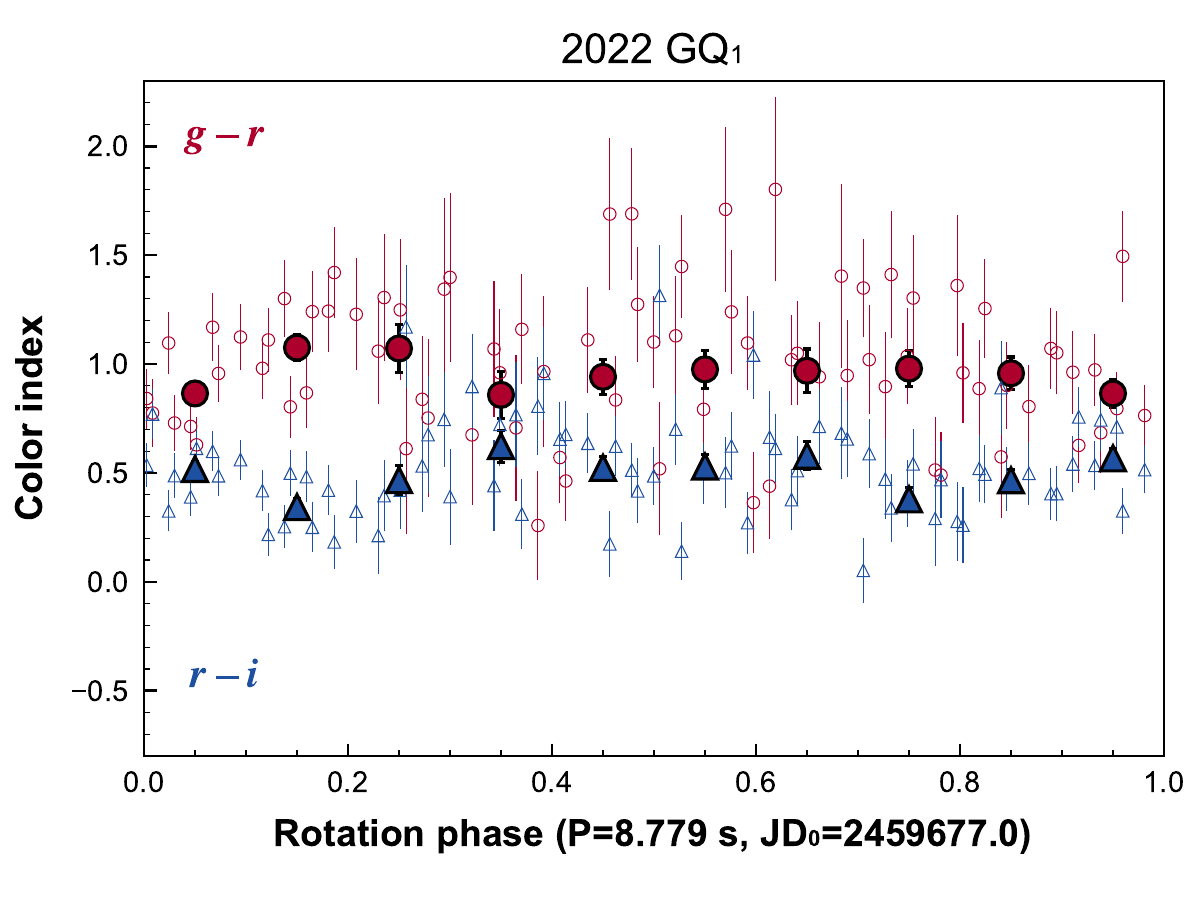}
\caption{
Same as Fig. \ref{fig:col_TY}, but for \GQ.
Phase zero is set to 
JD 2459505.2 (2022 April 07 12:00:00 UT).
}
\label{fig:col_GQ}
\end{figure}

To investigate possible surface heterogeneity of the observed NEAs,
we constructed rotational color time series as shown in Figs.~\ref{fig:col_TY}--\ref{fig:col_GQ}.
For each NEA, the data were folded using the derived rotation period and divided into 10 equal phase bins.
Within each bin, we computed the weighted mean color and its uncertainty.
The statistical significance of the variations was assessed using a weighted one-way analysis of variance (ANOVA) across phase bins, adopting a significance level of $p = 0.05$.
Each measurement $y_i$ with uncertainty $\sigma_i$ is assigned a weight given by the inverse variances of each color measurement, $w_i = 1/{\sigma_i^2}$. 
The weighted mean of bin $b$ and the grand mean are given by
\begin{eqnarray}
\bar{y}_b = \frac{\sum_{i \in b} w_i y_i}{\sum_{i \in b} w_i}, \\
\bar{y} = \frac{\sum_b \sum_{i \in b} w_i y_i}{\sum_b \sum_{i \in b} w_i}.
\end{eqnarray}
The weighted sums of squares between bins and within bins are expressed by
\begin{eqnarray}
SS_{\mathrm{between}} = \sum_b \left(\sum_{i \in b} w_i \right) (\bar{y}_b - \bar{y})^2, \\
SS_{\mathrm{within}} = \sum_b \sum_{i \in b} w_i (y_i - \bar{y}_b)^2,
\end{eqnarray}
and the $F$ statistic is
\begin{equation}
F = \frac{SS_{\mathrm{between}} / (k-1)}{SS_{\mathrm{within}} / (N-k)},
\end{equation}
where $k$ is the number of bins and $N$ is the total number of measurements.  
This $F$ quantifies the ratio of variance between bins to variance within bins.
The resulting $F$-value follows an $F$-distribution with $(d_1, d_2)$ degrees of freedom under the null hypothesis that the color is constant across phase bins, where $d_1 = k - 1$ and $d_2 = N - k$.

The resuls of ANOVA are summarized in Table \ref{tab:ANOVA}.
For \TY, both the $g-r$ and $r-i$ variations are statistically significant 
($p = 6.16\times10^{-8}$ and $p = 0.031$).
For \UW, both $g-r$ and $r-i$ are consistent with a constant color 
($p = 0.968$ and $0.958$, respectively).  
For \GQ, both $g-r$ and $r-i$ are also consistent with a constant color ($p = 0.693$ and $0.054$, respectively).  
Overall, statistically significant rotational color variation is detected only in the 
$g-r$ and $r-i$ colors
of \TY.

To quantitatively estimate the fraction of possible surface heterogeneity,
we assume a simple two-component model consisting of a "main" surface and a "spot" with a different color.
Let $C_{\rm main}$ and $C_{\rm spot}$ denote the intrinsic colors of
the main body and the anomalous patch, respectively.
Under the assumption of linear mixing,
the observed color at a given rotation phase, $C_{\rm obs}$, is determined by the projected area ratio of the spot, $f_{\rm spot}$, as follows:
\begin{equation}
    C_{\rm obs}(f_{\rm spot}) = (1 - f_{\rm spot}) C_{\rm main} + f_{\rm spot} C_{\rm spot}
\end{equation}
The maximum color variation expected in a rotational color time series, $\Delta C_{\rm obs, max}$,
corresponds to the difference between the phase where the spot has the maximum visibility
and the phase where the surface is dominated by the main component and the spot is invisible.
Thus, $\Delta C_{\rm obs, max}$ is expressed as:
\begin{align}
    \Delta C_{\rm obs, max} &= |C_{\rm obs}(f_{\rm spot, max}) - C_{\rm obs}(0)| \nonumber \\
                            &= f_{\rm spot, max} |C_{\rm main} - C_{\rm spot}|,
\end{align}
where $f_{\rm spot, max}$ is the maximum projected area ratio of the spot.
Rearranging this for $f_{\rm spot, max}$, we obtain:
\begin{equation}
    f_{\rm spot, max} = \frac{\Delta C_{\rm obs, max}}{|C_{\rm main} - C_{\rm spot}|}.
\end{equation}
It should be noted that $f_{\rm spot}$ represents the ratio of the projected area of the spot relative to the total projected area of the asteroid at the observed geometry, rather than its true physical surface area.
Conversion to the actual surface area would require further modeling of the asteroid's three-dimensional shape.

We define the maximum color difference, $\Delta C$, as the difference between the maximum and minimum binned colors, $\Delta C = |C_{\rm max} - C_{\rm min}|$, where $C$ can be either $g-r$ or $r-i$.
The resulting values are as follows:
$\Delta C_{g-r} = $\dCgrTY and $\Delta C_{r-i} =$ \dCriTY  for \TY,
$\Delta C_{g-r} = $\dCgrUW and $\Delta C_{r-i} =$ \dCriUW  for \UW,
and
$\Delta C_{g-r} = $\dCgrGQ  and $\Delta C_{r-i} =$ \dCriGQ for \GQ.

To assess statistical significance, we perform a hypothesis test against the null hypothesis that binned colors follow a normal distribution, $C \sim \mathcal{N}(\mu_C, \sigma_C^2)$, where $\mu_C$ and $\sigma_C$ are the mean and the standard deviation of $C$.
We define the normalized variables as $x = (C - \mu_C) / \sigma_C$, such that each $x$ follows a standard normal distribution $\mathcal{N}(0, 1)$. 
The normalized range, defined as $z = x_{\max} - x_{\min} = \Delta C / \sigma_C$, is invariant under a constant shift of the baseline $\mu_C$. 
The probability density function (PDF) of this range $z$ for a set of $k$ independent and identically distributed variables follows the general form:
\begin{align}
    \mathrm{pdf}(z) &= \nonumber \\
    \hspace*{-3em} &\frac{k!}{(k-2)!} \int_{-\infty}^{\infty} \phi(\xi) \phi(\xi - z) \left[ \Phi(\xi) - \Phi(\xi - z) \right]^{k-2} \mathrm{d}\xi
\end{align}
where $\phi$ and $\Phi$ are the PDF and cumulative distribution function (CDF) of the standard normal distribution, respectively. 
In this study, we adopt $k=10$ bins, 
which gives a factor of $k!/(k-2)! = 90$ accounting for the permutations of the two boundary values (maximum and minimum), and an exponent $k-2=8$ representing the remaining data points constrained within this interval.

We define the $3\sigma$ upper limit based on the 99.73\% percentile 
of the null hypothesis distribution for the normalized range $z = \Delta C / \sigma_C$. 
Numerical integration of the PDF derived above yields a threshold of $z < 5.64$. 
We note that a rigorous error propagation cannot be applied to $\Delta C$ 
because it is the maximum-minus-minimum of 10 independent bins. 
Instead, for simplicity, we assign $\sigma_C$ as the arithmetic mean of the uncertainties of $C_{\rm max}$ and $C_{\rm min}$.
Consequently, the statistical upper limits for the color variations in our samples are determined as follows:
$\Delta C_{g-r} < $\dCgrupperTY and $\Delta C_{r-i} < $ \dCriupperTY for \TY,
$\Delta C_{g-r} < $\dCgrupperUW and $\Delta C_{r-i} < $ \dCriupperUW for \UW,
and
$\Delta C_{g-r} < $\dCgrupperGQ  and $\Delta C_{r-i} < $ \dCriupperGQ  for \GQ.
For majority of observed NEAs, no color variation exceeds the $3\sigma$ significance threshold;
\TY exhibits a $g-r$ color variation that exceeds this level.

While the ANOVA test indicates statistically significant rotational color 
variations in the $g-r$ and $r-i$ colors
of \TY, 
the peak-to-peak amplitude estimated from $\Delta C$ does not exceed the 3$\sigma$ significance level
for the $r-i$ color.
This discrepancy suggests that although a small-amplitude systematic color change may be present, 
it is not sufficiently prominent to be distinguished from statistical fluctuations at a high significance level. 
For the other objects and colors, the observed $\Delta C$ values are entirely consistent with statistical noise.
For \UW, this is somewhat expected since the exposure time corresponds to approximately 25\% of the derived rotation period. 
Any existing surface heterogeneity would be difficult to detect, as the resulting lightcurve amplitude is smeared out by the long exposure.

\begin{figure}
\centering
\includegraphics[width=1.0\hsize]{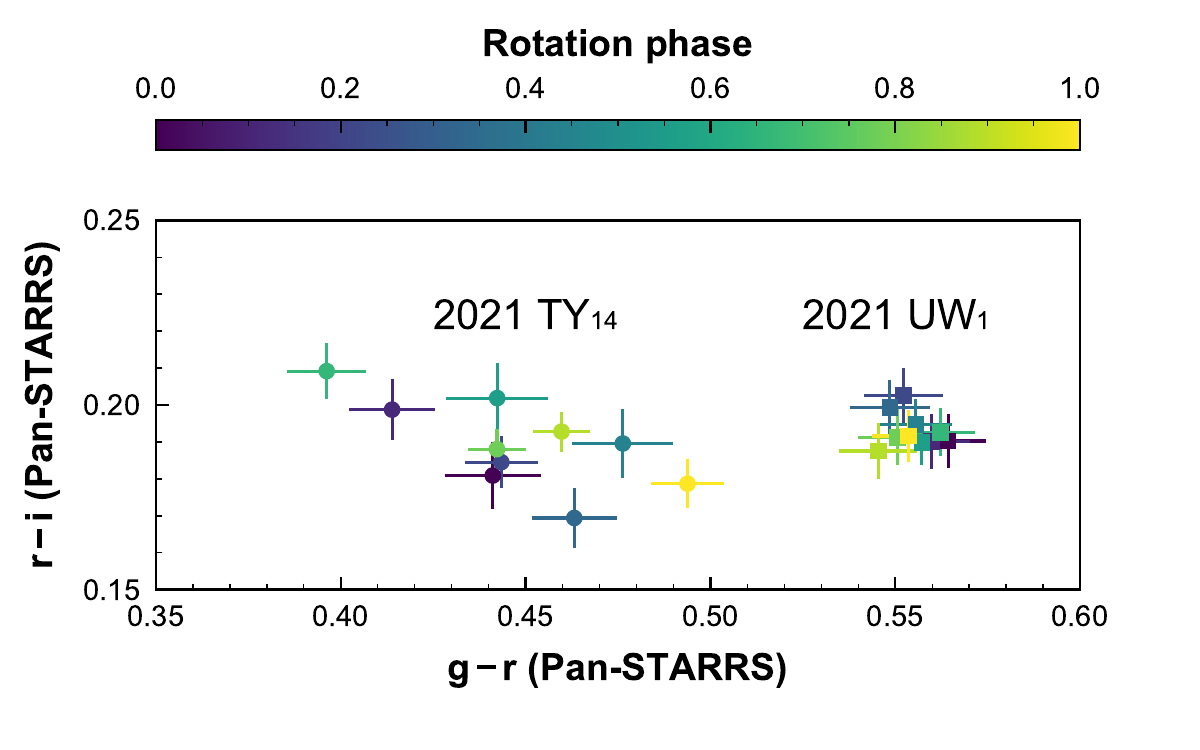}
\caption{
Distributions of binned colors on the $g-r$ vs. $r-i$ color-color diagram for \TY (circles) and \UW (squares). 
The marker colors indicate the rotation phase according to the color bar. Each data point and its associated error bars represent the weighted mean and its standard error for each phase bin, respectively.
}
\label{fig:cc_phase}
\end{figure}

Fig.~\ref{fig:cc_phase} presents the binned colors for each rotation phase on a color-color diagram. For \TY, the scatter exceeds the observational uncertainties, whereas for \UW, the data points remain well within the uncertainties. 
While these results are consistent with our analyses above, the lack of a monotonic trend for \TY in Fig.~\ref{fig:cc_phase} despite the expectation from a single-spot model suggests that the scatter may not be physical in origin or cannot be explained by our current model assuming only a single spot.

Assuming a spot that is 0.2~mag redder (e.g., an S-complex–like patch on an X-complex surface; see Fig.~\ref{fig:cc}), we estimate the projected spot fractions, $f_{\rm spot, max}$, based on the observed color variations. For \TY, the observed $g-r$ variation of $\Delta C = 0.098$~mag is statistically significant, yielding a maximum spot fraction of $f_{\rm spot, max} = 0.49$. 
For other cases where no significant variation was detected, we use the $3\sigma$ upper limits on $\Delta C$ to conservatively estimate the corresponding upper limits on $f_{\rm spot, max}$. 
We find $f_{\rm spot,max} = \fspotmaxTY$ for \TY with 
the observed $r-i$ variation.
Adopting the stronger constraint from $r-i$ variation,
we find $f_{\rm spot,max} = \fspotmaxUW$ for \UW.
For \GQ, the constraints are weaker, and we cannot place a meaningful upper limit on $f_{\rm spot,max}$.
These results indicate that the surfaces of the observed NEAs are largely homogeneous, although the statistically significant $g-r$ variation detected for \TY\ suggests a heterogeneous surface with a projected spot fraction of approximately 50\%. For the other cases, minor localized heterogeneity covering up to $\sim$20\% of the visible surface cannot be ruled out.
The adopted contrast of 0.2~mag is somewhat arbitrary, as the expected color variation for such small asteroids is not well constrained; for example, observations of the \textit{Hayabusa2} target Ryugu ($D\sim900$~m) show surface color variability at the level of $\sim$0.05~mag \citep[][their Figure~14]{Tatsumi2020}, in which case detecting localized patches would be more challenging.

Recently, \citet{Hasegawa2024} investigated candidates for surface heterogeneity in a sample of 130 main-belt asteroids using multiepoch visible-to-near-infrared spectroscopic data from the MIT–Hawaii Near-Earth Object Spectroscopic Survey using
the NASA IRTF.
They showed that surface heterogeneity is not limited to large asteroids, identifying candidates for such features in much smaller objects, with diameters down to approximately 5~km.
Surface heterogeneity has also been reported in even smaller NEAs \citep[e.g., ][]{Shepard2008, Lopez-Oquendo2022, Devogele2024a}.
This study provided the first evidence of surface heterogeneity in tiny asteroids smaller than 100~m, suggesting that many tiny NEAs may be similar to 2008~TC$_3$, which was recovered as the Almahata Sitta meteorite and found to be a mixture of different meteorite types \citep{Jenniskens2009, Jenniskens2010}.
Simultaneous multicolor photometry of fast-rotating asteroids has been demonstrated to be an effective method for probing surface heterogeneity.
Ongoing and future space missions targeting small asteroids, such as the \textit{Hayabusa2} extended mission for 1998~KY$_{26}$ \citep{Hirabayashi2021, Kikuchi2023} and the \textit{Tianwen-2} mission for (469219)~Kamo`oalewa \citep{Zhang2021}, will provide comparative information on their surface homogeneity.

\begin{deluxetable*}{llcccccc}
\tablenum{3}
\tablecaption{Results of the ANOVA tests for rotational color variations.\label{tab:ANOVA}}
\tablewidth{0pt}
\tablehead{
Object & Color & $d_1$ & $d_2$ & $SS_{\rm between}$ & $SS_{\rm within}$ & $F$ & $p$ 
}
\decimals
\startdata
  \TY & $g-r$ & 9 & 742 & 63.45 & 890.87 & 5.87 & $6.16\times10^{-8}$ \\
  & $r-i$ & 9 & 742 & 20.23 & 808.51 & 2.06 & 0.031 \\
\UW  & $g-r$ & 9 & 515 & 3.09 & 549.56 & 0.32 & 0.968 \\
 & $r-i$ & 9 & 515 & 3.54 & 579.19& 0.35 & 0.958 \\
\GQ  & $g-r$ & 9 & 70 & 12.13 & 131.92 & 0.72 & 0.693 \\
 & $r-i$ & 9 & 70 & 28.88 & 113.37 & 1.98 & 0.054 \\
\enddata
\end{deluxetable*}

\subsection{Diameter-period relation with taxonomic complexes} \label{subsec:DP}
\begin{figure*}
\centering
\includegraphics[width=1.0\hsize]{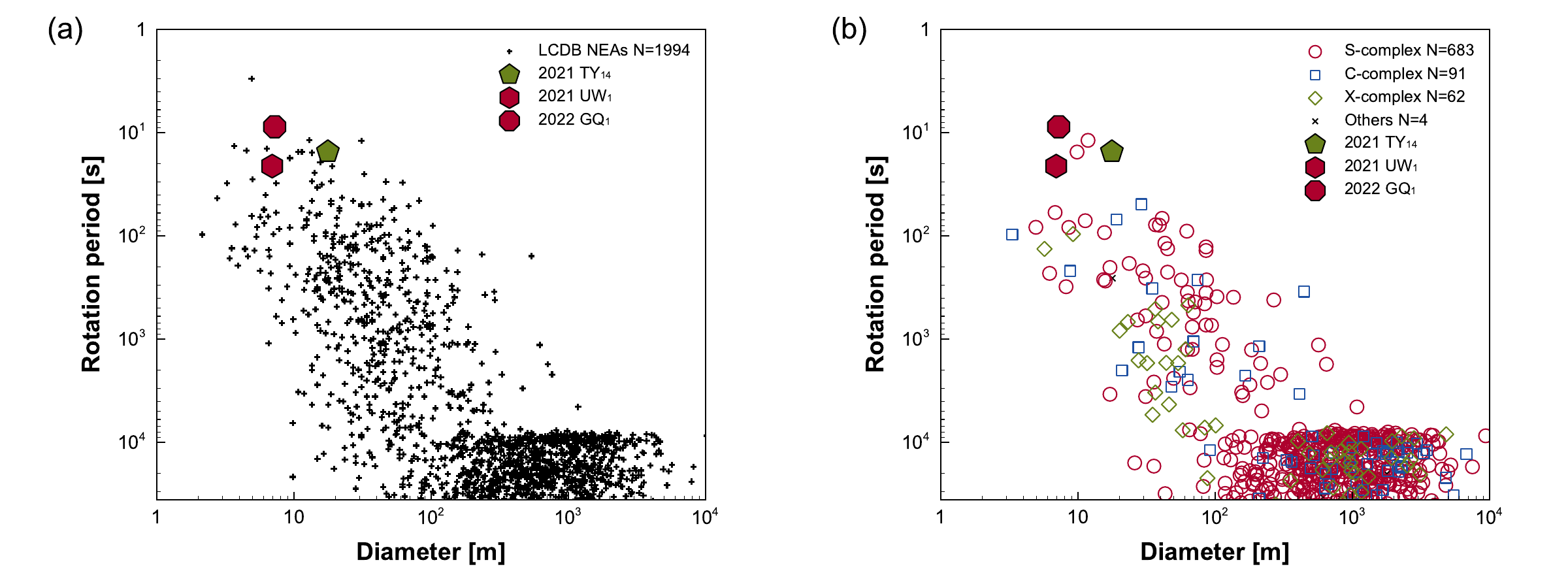}
\caption{
(a) Diameter–rotation period relations for objects in the LCDB \citep{Warner2009} as of October 2023 are shown as plus signs.
The three NEAs observed in this work are shown 
as a pentagon (\TY), a hexagon (\UW), and an octagon (\GQ).
(b) Diameter–rotation period relations for objects with taxonomic complexes 
in the literature (see the main text for details).
S-complex, C-complex, X-complex, and other types are shown with circles, squares, diamonds, and crosses, respectively.
Diameter is derived from $H$ assuming geometric albedo in V-band, $p_\mathrm{V}$, as follows:
$p_\mathrm{V}=0.20$ for S-complex, 
$p_\mathrm{V}=0.07$ for C-complex, 
$p_\mathrm{V}=0.10$ for X-complex, 
and $p_\mathrm{V}=0.168$ for others.
}
\label{fig:DP}
\end{figure*}

The diameter--rotation period diagram (Fig. \ref{fig:DP}) could tell us the dynamical history of small bodies \citep[e.g.,][]{Holsapple2007, Kwiatkowski2010c, Sanchez2014, Thirouin2016, Thirouin2018, Beniyama2022}.
For larger bodies (typically $D \geq 200$~m), 
it has been established that they are constrained by the cohesionless spin barrier, 
and a dependence on spectral type has been reported and discussed \citep[][]{Perna2016, Carbognani2017, Rondon2020}.
In contrast, smaller objects can rotate faster than the cohesionless spin barrier and are expected to deviate from this limit due to material strength. 
However, observational constraints remain limited.

We estimated both the taxonomic complexes and rotation periods of three sub-minute rotators through multicolor video observations, as shown in panel (b) of Fig.~\ref{fig:DP}.
We compared our results with literature data from the LCDB \citep{Warner2009}, 
using only NEAs with reliable period estimates (the quality code $U \geq 2$), and cross-matched them with published taxonomic classifications
\citep{Binzel2004, Binzel2019, Devogele2019, Mommert2016, Marsset2022a, Sergeyev2021, Sanchez2024}. 
For consistency, all reported spectral types, originally derived using different taxonomies, have been grouped into S-, C-, and X-complexes, with remaining types categorized as others.
The identification of spectrally classified objects in this extreme rotational regime provides valuable constraints on the internal structure and material strength of small asteroids \citep{Holsapple2004, Holsapple2007}.
A detailed statistical analysis of taxonomic complex trends in the diameter–period diagram is beyond the scope of this study.

As \citet{Greenstreet2026} demonstrated, many fast-rotating asteroids are expected to be discovered by
the Rubin Observatory Legacy Survey of Space and Time (LSST)
over the next decade.
However, even with Rubin, detecting sub-minute rotators whose rotation periods are comparable to typical exposure times will be challenging, although such objects may still be detectable through streak photometry \citep[e.g,][]{Bolin2024, Devogele2024b}.
Therefore, video observations like those presented in this study will be crucial for the physical characterization of 
asteroids with sub-minute rotation periods.
Furthermore, the simultaneous determination of rotation periods and taxonomic types is crucial for planetary defense. 
These properties provide essential constraints on the internal structure and composition of tiny NEAs, which are necessary for assessing potential impact damage. 
Our simultaneous multiband approach allows for the rapid characterization of Earth impactors 
discovered shortly before the impact. 
Given the extremely narrow observation windows for such objects \citep[e.g.,][]{Devogele2024b, Kareta2024b}, 
the ability to obtain both lightcurves and spectral information in a single, 
efficient observation run is a significant advantage for real-time hazard assessment.

\section{Conclusion}
We conducted simultaneous tricolor video observations of three tiny near-Earth asteroids, \TY, \UW, and \GQ.
We derived their rotation periods and visible colors, confirming that all three are ultra-fast-rotating asteroids with rotation periods shorter than 60~s.
The derived colors indicate that \TY\ is consistent with the X-complex, while \UW\ and \GQ\ are consistent with S-complex asteroids.
Analysis of the color variations over their rotation allows us to place constraints on surface heterogeneity. 
The results suggest that the surfaces of these asteroids are generally homogeneous; however, \TY exhibits statistically significant $g-r$ color heterogeneity, consistent with a projected spot fraction of approximately 50\%. 
For 2021~UW$_{1}$, minor localized variations of up to $\sim20$\% in composition may still be present.
Our simultaneous measurements effectively eliminated the color uncertainties
typically introduced by the rapid rotation and apparent motion of these tiny NEAs during sequential filter changes.
Although many fast-rotating asteroids are expected to be discovered by the Rubin Observatory in the next decade, detecting fast rotators with periods comparable to typical exposure times will remain challenging.
High-cadence, multicolor observations with small to medium ground-based telescopes will therefore remain essential for characterizing these tiny objects.
Such efforts are crucial not only for improving our understanding of their rotation states and surface properties but also for refining impact hazard assessments and supporting planetary defense strategies.
\clearpage
\nolinenumbers

\section*{Acknowledgment}
The authors thank the anonymous referee for their helpful comments and valuable suggestions.
This work was supported by JSPS KAKENHI Grant Numbers JP22K21344, JP23KJ0640, and 25H00665.
This work was supported by the French government through the France 2030
investment plan managed by the National Research Agency (ANR), as part of the
Initiative of Excellence Université Côte d’Azur under reference number ANR-15-IDEX-01.
This work was supported by the Japan Society for the Promotion of Science (JSPS) Overseas Research Fellowships.
This research is partially supported by the Optical
and Infrared Synergetic Telescopes for Education and Research
(OISTER) program funded by MEXT of Japan.
The authors thank the TriCCS developer team (which has been supported by the JSPS KAKENHI grant Nos. JP18H05223,
JP20H00174, and JP20H04736, and by NAOJ Joint Development Research).
The Pan-STARRS1 Surveys (PS1) and the PS1 public science archive have been
made possible through contributions by the Institute for Astronomy,
the University of Hawaii, the Pan-STARRS Project Office,
the Max-Planck Society and its participating institutes,
the Max Planck Institute for Astronomy, Heidelberg and
the Max Planck Institute for Extraterrestrial Physics, Garching,
The Johns Hopkins University, Durham University, the University of Edinburgh,
the Queen's University Belfast, the Harvard-Smithsonian Center for Astrophysics,
the Las Cumbres Observatory Global Telescope Network Incorporated,
the National Central University of Taiwan, the Space Telescope Science Institute,
the National Aeronautics and Space Administration under Grant No. NNX08AR22G
issued through the Planetary Science Division of the NASA Science Mission Directorate,
the National Science Foundation Grant No. AST-1238877, the University of Maryland,
Eotvos Lorand University (ELTE), the Los Alamos National Laboratory,
and the Gordon and Betty Moore Foundation.

\facilities{Seimei (TriCCS)}
\software{
    NumPy \citep{Oliphant2015, Harris2020},
    pandas \citep{Reback2021},
    SciPy \citep{Virtanen2020},
    AstroPy \citep{Astropy2013, Astropy2018},
    astroquery \citep{Ginsburg2019}
}

\appendix

\section{Lightcurves}
\begin{figure*}
\centering
\includegraphics[width=1.0\hsize]{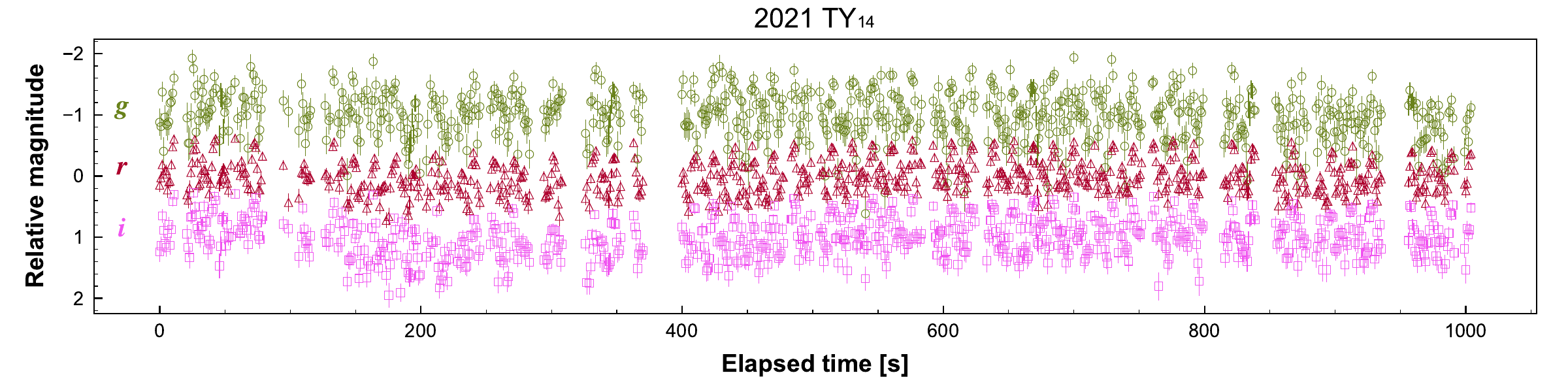}
\caption{
Lightcurves of \TY.
From top to bottom, the $g$, $r$, and $i$ bands are plotted using circles, triangles, and squares, respectively.
The $g$ and $i$ bands lightcurves are vertically offset by $-1$ and $+1$~mag, respectively.
Bars indicate the 1$\sigma$ photometric uncertainties. 
}
\label{fig:lc_full_TY}
\end{figure*}

\begin{figure*}
\centering
\includegraphics[width=1.0\hsize]{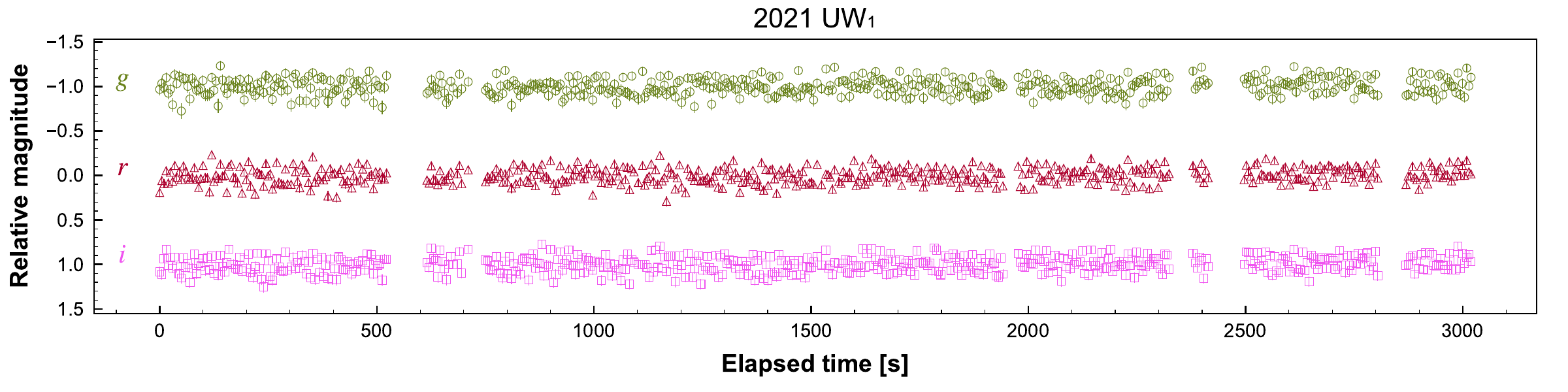}
\caption{
Same as Fig. \ref{fig:lc_full_TY}, but for
\UW.
}
\label{fig:lc_full_UW}
\end{figure*}

\begin{figure*}
\centering
\includegraphics[width=1.0\hsize]{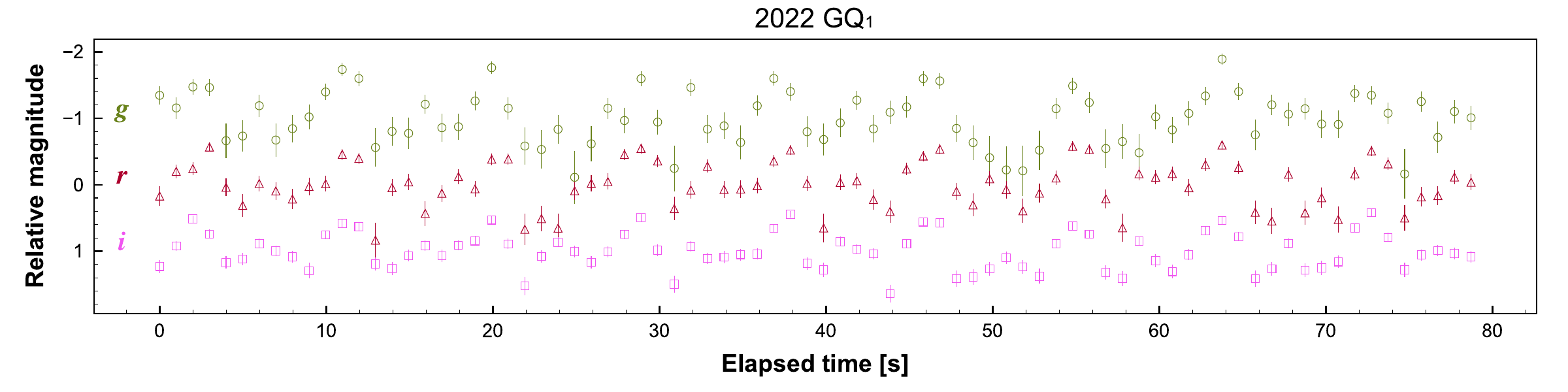}
\caption{
Same as Fig. \ref{fig:lc_full_TY}, but for
\GQ.
}
\label{fig:lc_full_GQ}
\end{figure*}

Lightcurves of the NEAs are presented here in Figs. \ref{fig:lc_full_TY}--\ref{fig:lc_full_GQ}.

\section{Periodic analysis}
\begin{figure*}[htbp]
\centering
\begin{minipage}[t]{0.3\hsize}
  \centering
  \includegraphics[width=\hsize]{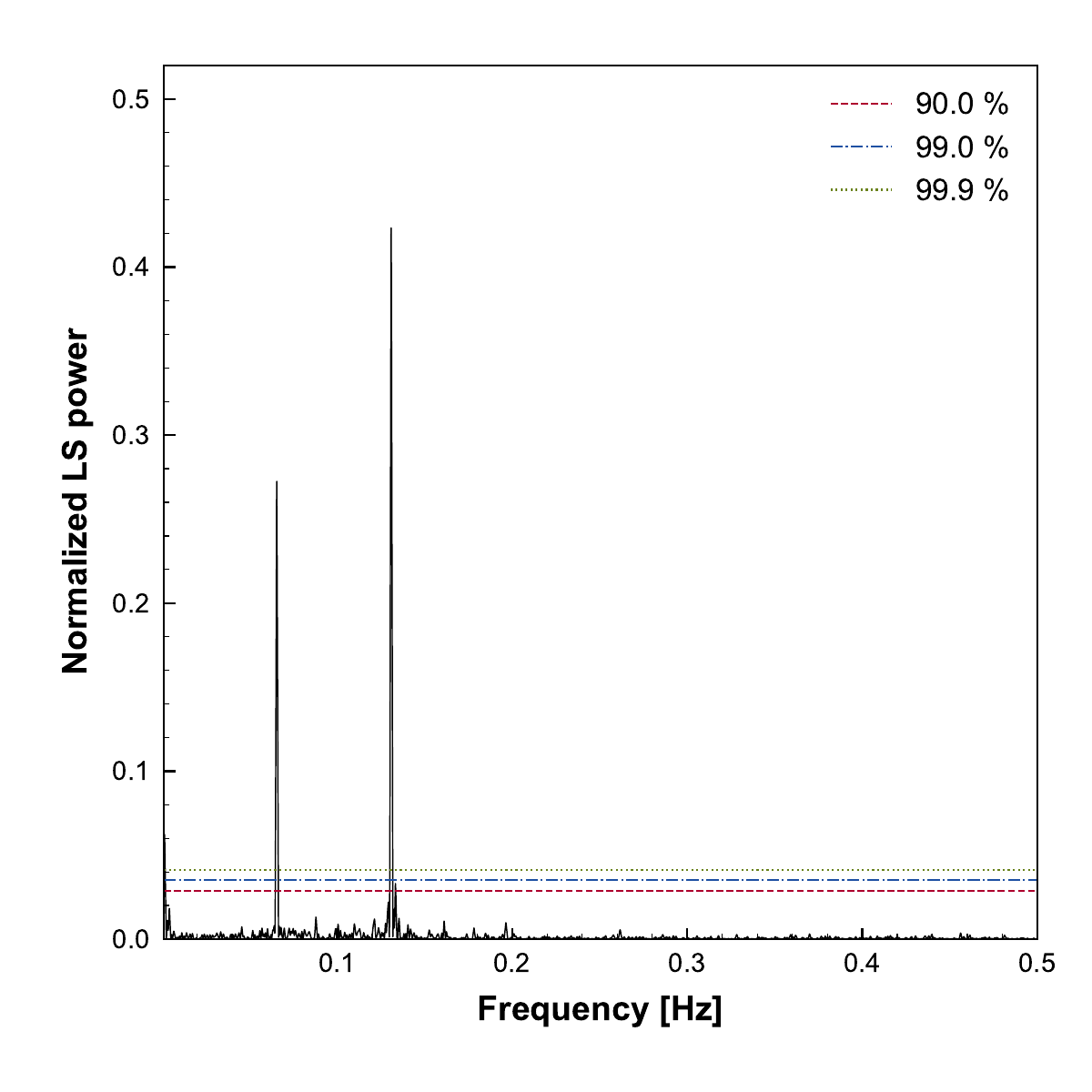}
  \caption{
    Lomb--Scargle periodogram of \TY.
    Horizontal dashed, dash-dotted, and dotted lines show 90.0, 99.0, and 99.9\% confidence levels, respectively.
    }
  \label{fig:LS_TY}
\end{minipage}
\hspace{0.02\hsize}
\begin{minipage}[t]{0.3\hsize}
  \centering
  \includegraphics[width=\hsize]{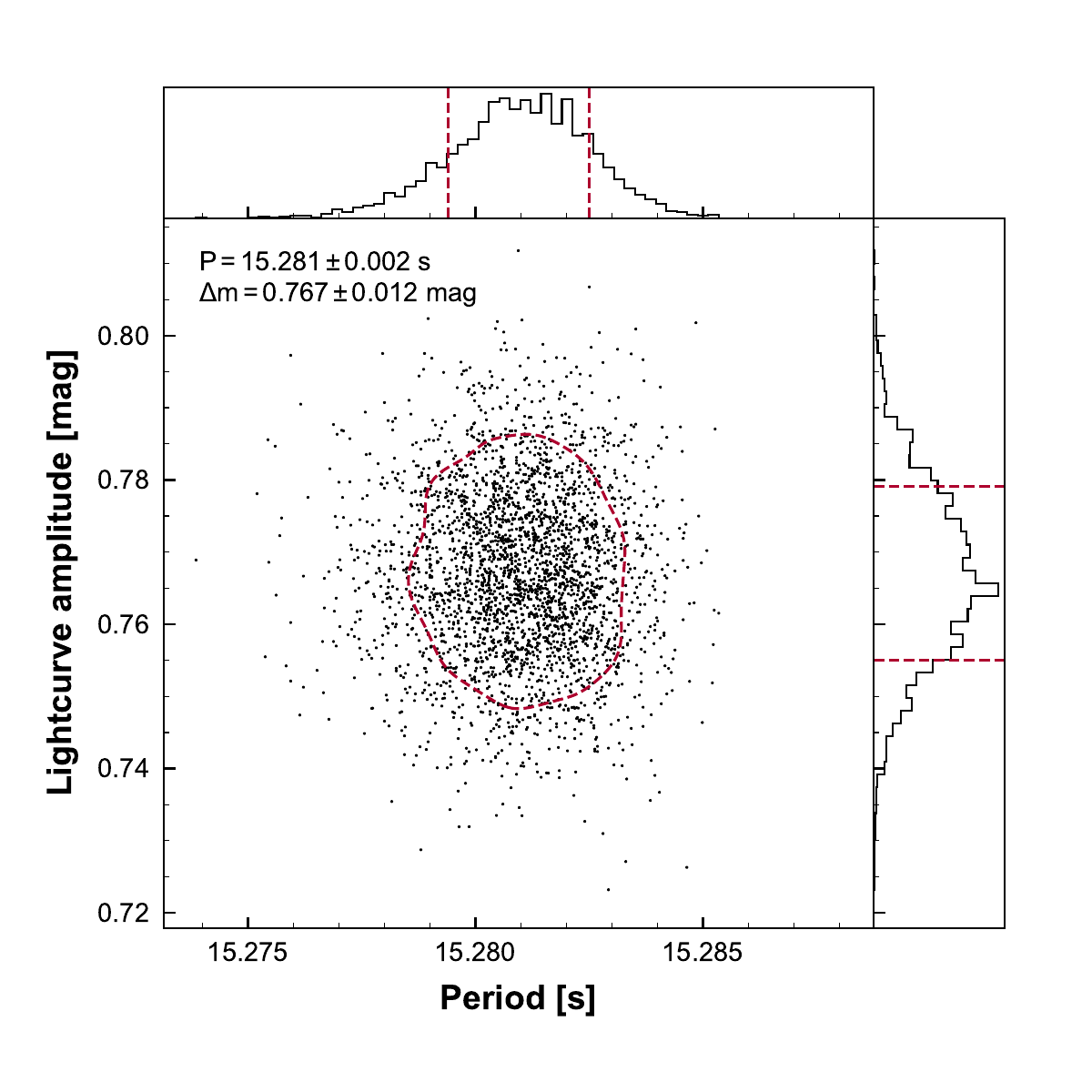}
  \caption{
  Scatter plot of the rotation periods and the light curve amplitudes of \Nmc model curves of \TY. 
  Histograms at the top and to the side present the marginal distributions of the periods and the amplitudes, respectively. 
  The corresponding 1$\sigma$ standard deviations are calculated from the marginal histograms of the period and amplitude, respectively.
  In the period–amplitude diagram, the dashed lines enclose the 1$\sigma$ confidence region in the period–amplitude space for the derived solutions.
}
  \label{fig:LS_MC_TY}
\end{minipage}
\end{figure*}

\begin{figure*}[htbp]
\centering
\begin{minipage}[t]{0.3\hsize}
  \centering
  \includegraphics[width=\hsize]{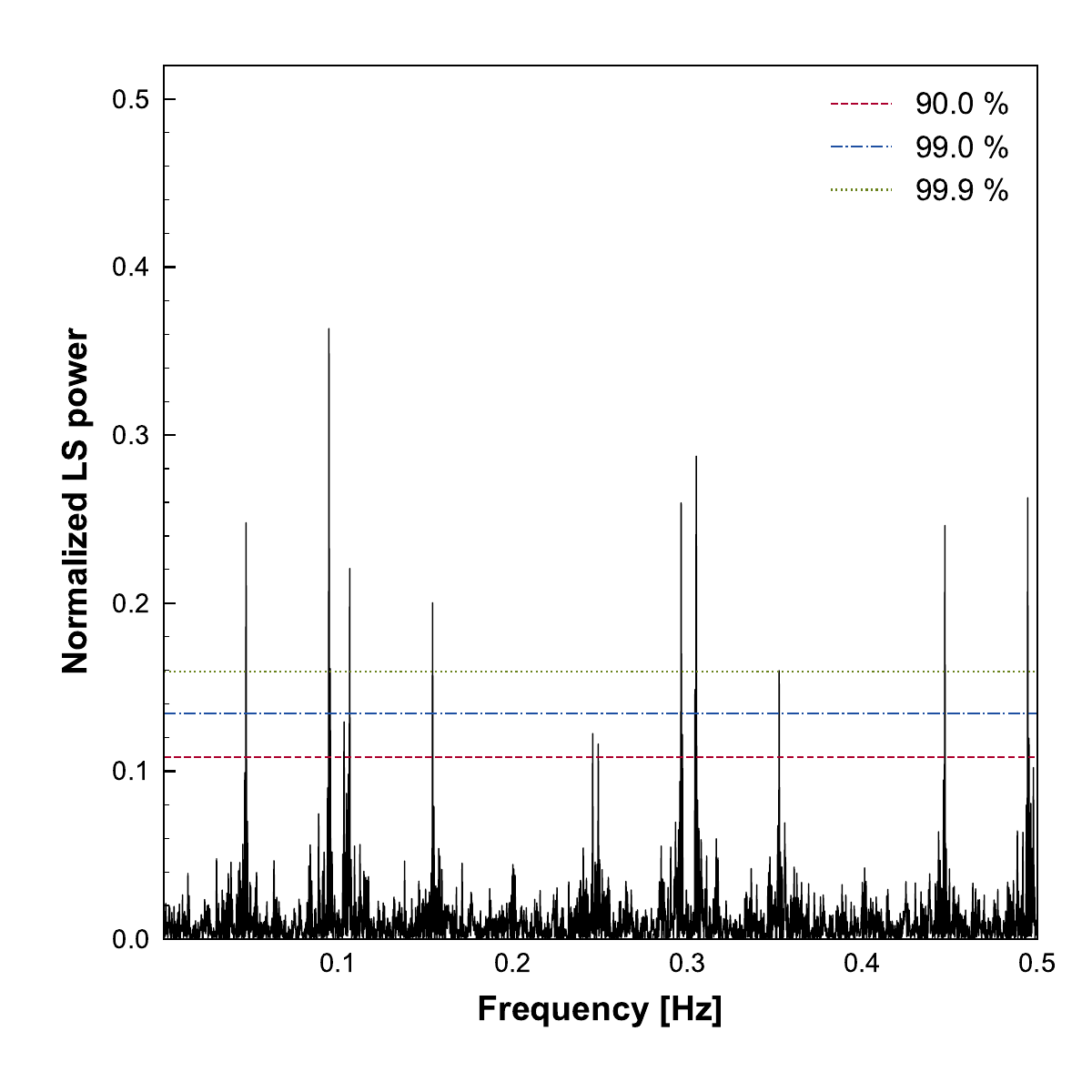}
  \caption{
  Same as Fig. \ref{fig:LS_TY}, but for \UW.
  }
  \label{fig:LS_UW}
\end{minipage}
\hspace{0.02\hsize}
\begin{minipage}[t]{0.3\hsize}
  \centering
  \includegraphics[width=\hsize]{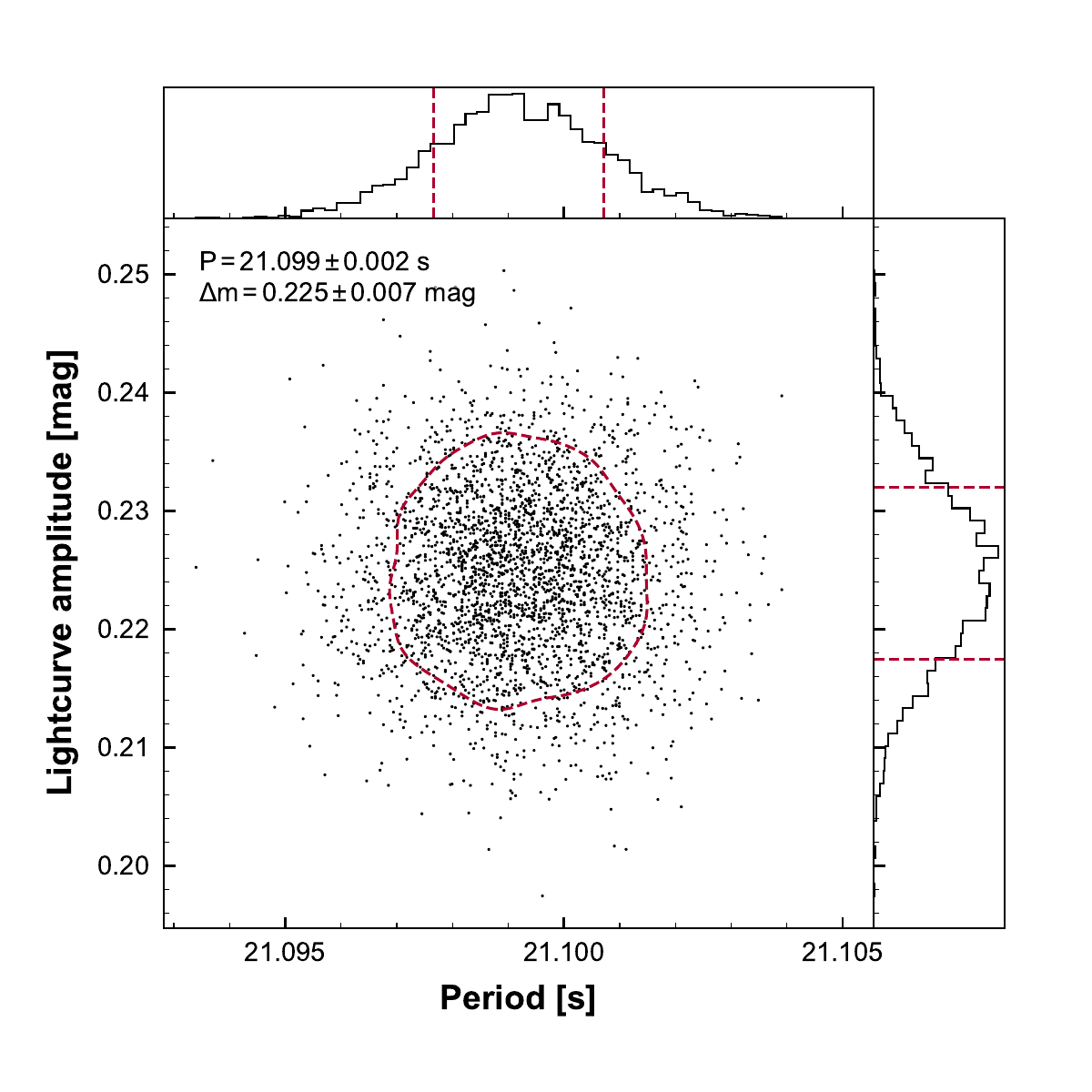}
  \caption{
    Same as Fig. \ref{fig:LS_MC_TY}, but for \UW.
    }
  \label{fig:LS_MC_UW}
\end{minipage}
\end{figure*}

\begin{figure*}[htbp]
\centering
\begin{minipage}[t]{0.3\hsize}
  \centering
  \includegraphics[width=\hsize]{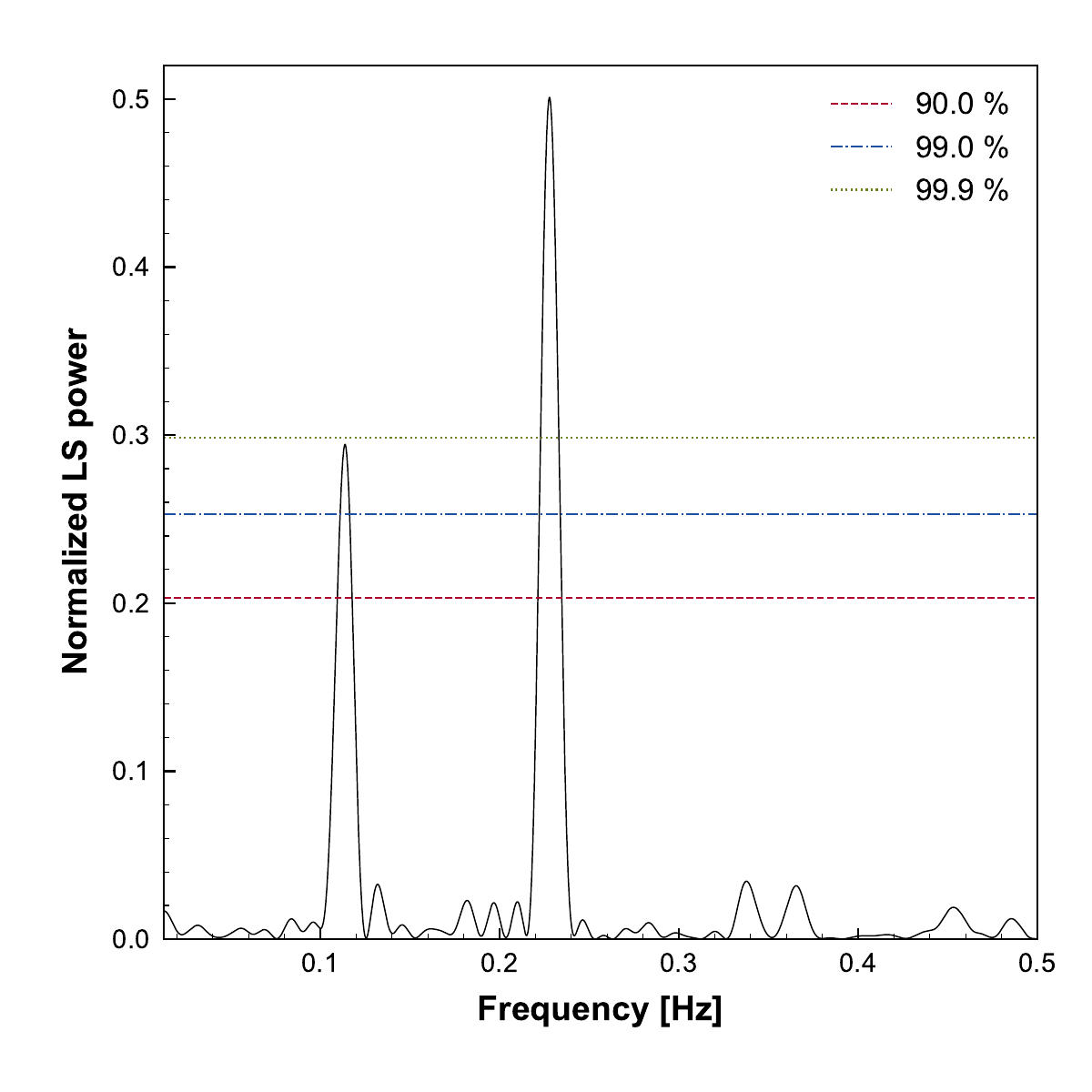}
  \caption{
  Same as Fig. \ref{fig:LS_TY}, but for \GQ.
  }
  \label{fig:LS_GQ}
\end{minipage}
\hspace{0.02\hsize}
\begin{minipage}[t]{0.3\hsize}
  \centering
  \includegraphics[width=\hsize]{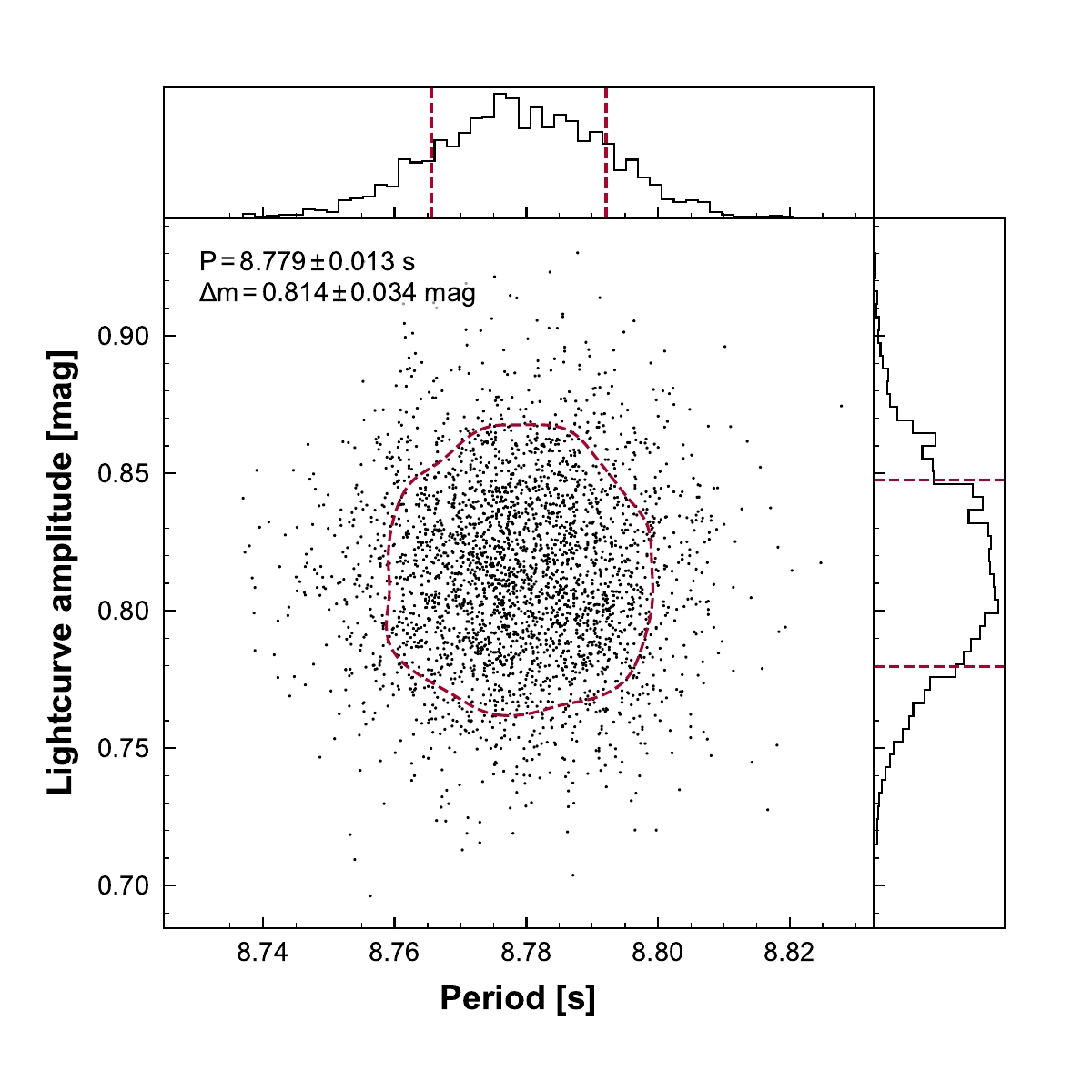}
  \caption{
    Same as Fig. \ref{fig:LS_MC_TY}, but for \GQ.
    }
  \label{fig:LS_MC_GQ}
\end{minipage}
\end{figure*}

Lomb--Scargle periodograms of the three NEAs are presented in 
Figs. \ref{fig:LS_TY}, \ref{fig:LS_UW}, and \ref{fig:LS_GQ}.
We showed 90.0, 99.0, and 99.9\% confidence levels in the periodogram.
The number of harmonics of the model curve was set to unity.
The uncertainties of the rotation period and lightcurve amplitude were estimated using a Monte Carlo method, in which the number of harmonics in the model curves was set to two.
For each NEA, we generated \Nmc synthetic lightcurves by randomly resampling the observed data, assuming that each data point follows a normal distribution with a standard deviation equal to its photometric error.
We then calculated the period and amplitude for each of the \Nmc lightcurves using the corresponding peak frequencies.
The standard deviations of these \Nmc\ measurements were adopted as the uncertainties, with the resulting distributions shown in Figs. \ref{fig:LS_MC_TY}, \ref{fig:LS_MC_UW}, and \ref{fig:LS_MC_GQ}."

\input {main.bbl}

\bibliographystyle{aasjournal}

\end{document}

%% file: tab_phot.tex
\begin{deluxetable*}{lrcccccccccc}
        \tablenum{1}
        \tablecaption{\label{t7}Summary of the observations\label{tab:obs}}
        \tablewidth{0pt}
        \tablehead{
        Object & Obs. Date & $t_\mathrm{exp}$ & $N_\mathrm{frame}$ & V     & $\alpha$ & $v$                & Air Mass   & Seeing   \\
               & (UTC)     & (s)              &                  & (mag) & (deg)    & (arcsec s$^{-1}$)  &            & (arcsec)
        }
        \decimals
        \startdata
        2021 TY$_{14}$&2021 Oct 15 14:42:17--14:59:02& 1 & 752 & 17.0 & 23.7 & 2.10 & 1.00--1.00 & 3.2\\
2021 UW$_1$&2021 Oct 29 11:02:22--11:52:41& 5 & 525 & 17.6 & 24.5 & 0.42 & 1.10--1.20 & 3.5\\
2022 GQ$_1$&2022 Apr 07 11:04:31--11:05:50& 1 & 80 & 16.7 & 81.3 & 7.35 & 1.62--1.62 & 2.5\\
\enddata \tablecomments{
            Observation time in UT in midtime of exposure (Obs. Date),
            exposure time per frame ($t_{\mathrm{exp}}$),
            and the number of frames ($N_\mathrm{frame}$) are listed.
            Predicted V band apparent magnitude (V), 
            phase angle ($\alpha$),
            and 
            apparent angular rate of asteroids ($v$)
            at the observation starting time
            are referred to NASA Jet Propulsion Laboratory (JPL) Horizons
             as of May 10, 2026.
            Elevations of asteroids to calculate air mass range (Air Mass) are 
            also referred to NASA JPL Horizons.
            Seeing FWHM (Seeing) in the $r$ band for Seimei observations
            measured by computing the FWHM of reference stars are also listed.
            }
            \end{deluxetable*}

%% file: tab_res.tex
\begin{deluxetable*}{ccccccc}
        \tablenum{2}
        \tablecaption{\label{t7}Summary of observational results\label{tab:res}}
        \tablewidth{0pt}
        \tablehead{
        Object & $H$  & $P$   & $\Delta m$  & $g-r$ & $r-i$ & Taxonomic complex \\
               & (mag)& (s)   & (mag)       & (mag) & (mag) &                   
        }
        \decimals
        \startdata
        2021 TY$_{14}$ & 27.27 & $15.281\pm0.002$ & $0.767\pm0.012$ & $0.448\pm0.003$ & $0.189\pm0.002$ &X\\
2021 UW$_1$ & 26.16 & $21.099\pm0.002$ & $0.225\pm0.007$ & $0.555\pm0.003$ & $0.193\pm0.002$ &S\\
2022 GQ$_1$ & 28.07 & $8.779\pm0.013$ & $0.814\pm0.034$ & $0.562\pm0.040$ & $0.182\pm0.031$ &S\\
\enddata
            \tablecomments{
            Absolute magnitudes ($H$)
            are referred to NASA Jet Propulsion Laboratory (JPL) Horizons
             as of May 14, 2026.
            The uncertainties following the color values represent statistical errors, while the typical systematic uncertainty in the color calibration is estimated to be $\sim$0.01 mag (see the main text for details).
            }
            \end{deluxetable*}

%% file: main.bbl
 \newcommand{\noopsort}[1]{}

%% file: main_AJ.bbl
\begin{thebibliography}{}
\expandafter\ifx\csname natexlab\endcsname\relax\def\natexlab#1{#1}\fi
\providecommand{\url}[1]{\href{#1}{#1}}
\providecommand{\dodoi}[1]{doi:~\href{http://doi.org/#1}{\nolinkurl{#1}}}
\providecommand{\doeprint}[1]{\href{http://ascl.net/#1}{\nolinkurl{http://ascl.net/#1}}}
\providecommand{\doarXiv}[1]{\href{https://arxiv.org/abs/#1}{\nolinkurl{https://arxiv.org/abs/#1}}}

\bibitem[{{Alarcon} {et~al.}(2026{\natexlab{a}}){Alarcon}, {Licandro}, \& {Serra-Ricart}}]{Alarcon2026}
{Alarcon}, M.~R., {Licandro}, J., \& {Serra-Ricart}, M. 2026{\natexlab{a}}, \aap, 707, A262, \dodoi{10.1051/0004-6361/202558528}

\bibitem[{{Alarcon} {et~al.}(2026{\natexlab{b}}){Alarcon}, {Licandro}, {Serra-Ricart}, {Garcia-{\'A}lvarez}, \& {Cabrera-Lavers}}]{Alarcon2026b}
{Alarcon}, M.~R., {Licandro}, J., {Serra-Ricart}, M., {Garcia-{\'A}lvarez}, D., \& {Cabrera-Lavers}, A. 2026{\natexlab{b}}, arXiv e-prints, arXiv:2605.04784, \dodoi{10.48550/arXiv.2605.04784}

\bibitem[{{Astropy Collaboration} {et~al.}(2013){Astropy Collaboration}, {Robitaille}, {Tollerud}, {Greenfield}, {Droettboom}, {Bray}, {Aldcroft}, {Davis}, {Ginsburg}, {Price-Whelan}, {Kerzendorf}, {Conley}, {Crighton}, {Barbary}, {Muna}, {Ferguson}, {Grollier}, {Parikh}, {Nair}, {Unther}, {Deil}, {Woillez}, {Conseil}, {Kramer}, {Turner}, {Singer}, {Fox}, {Weaver}, {Zabalza}, {Edwards}, {Azalee Bostroem}, {Burke}, {Casey}, {Crawford}, {Dencheva}, {Ely}, {Jenness}, {Labrie}, {Lim}, {Pierfederici}, {Pontzen}, {Ptak}, {Refsdal}, {Servillat}, \& {Streicher}}]{Astropy2013}
{Astropy Collaboration}, {Robitaille}, T.~P., {Tollerud}, E.~J., {et~al.} 2013, \aap, 558, A33, \dodoi{10.1051/0004-6361/201322068}

\bibitem[{{Astropy Collaboration} {et~al.}(2018){Astropy Collaboration}, {Price-Whelan}, {Sip{\H{o}}cz}, {G{\"u}nther}, {Lim}, {Crawford}, {Conseil}, {Shupe}, {Craig}, {Dencheva}, {Ginsburg}, {VanderPlas}, {Bradley}, {P{\'e}rez-Su{\'a}rez}, {de Val-Borro}, {Aldcroft}, {Cruz}, {Robitaille}, {Tollerud}, {Ardelean}, {Babej}, {Bach}, {Bachetti}, {Bakanov}, {Bamford}, {Barentsen}, {Barmby}, {Baumbach}, {Berry}, {Biscani}, {Boquien}, {Bostroem}, {Bouma}, {Brammer}, {Bray}, {Breytenbach}, {Buddelmeijer}, {Burke}, {Calderone}, {Cano Rodr{\'\i}guez}, {Cara}, {Cardoso}, {Cheedella}, {Copin}, {Corrales}, {Crichton}, {D'Avella}, {Deil}, {Depagne}, {Dietrich}, {Donath}, {Droettboom}, {Earl}, {Erben}, {Fabbro}, {Ferreira}, {Finethy}, {Fox}, {Garrison}, {Gibbons}, {Goldstein}, {Gommers}, {Greco}, {Greenfield}, {Groener}, {Grollier}, {Hagen}, {Hirst}, {Homeier}, {Horton}, {Hosseinzadeh}, {Hu}, {Hunkeler}, {Ivezi{\'c}}, {Jain}, {Jenness}, {Kanarek}, {Kendrew}, {Kern}, {Kerzendorf}, {Khvalko}, {King}, {Kirkby}, {Kulkarni},
  {Kumar}, {Lee}, {Lenz}, {Littlefair}, {Ma}, {Macleod}, {Mastropietro}, {McCully}, {Montagnac}, {Morris}, {Mueller}, {Mumford}, {Muna}, {Murphy}, {Nelson}, {Nguyen}, {Ninan}, {N{\"o}the}, {Ogaz}, {Oh}, {Parejko}, {Parley}, {Pascual}, {Patil}, {Patil}, {Plunkett}, {Prochaska}, {Rastogi}, {Reddy Janga}, {Sabater}, {Sakurikar}, {Seifert}, {Sherbert}, {Sherwood-Taylor}, {Shih}, {Sick}, {Silbiger}, {Singanamalla}, {Singer}, {Sladen}, {Sooley}, {Sornarajah}, {Streicher}, {Teuben}, {Thomas}, {Tremblay}, {Turner}, {Terr{\'o}n}, {van Kerkwijk}, {de la Vega}, {Watkins}, {Weaver}, {Whitmore}, {Woillez}, {Zabalza}, \& {Astropy Contributors}}]{Astropy2018}
{Astropy Collaboration}, {Price-Whelan}, A.~M., {Sip{\H{o}}cz}, B.~M., {et~al.} 2018, \aj, 156, 123, \dodoi{10.3847/1538-3881/aabc4f}

\bibitem[{{Beniyama}(2025)}]{Beniyama2025c}
{Beniyama}, J. 2025, \pasj, 77, L71, \dodoi{10.1093/pasj/psaf097}

\bibitem[{{Beniyama} {et~al.}(2025){Beniyama}, {Bolin}, {Sergeyev}, {Delbo}, {Abron}, {Belyakov}, {Sekiguchi}, \& {Takagi}}]{Beniyama2025b}
{Beniyama}, J., {Bolin}, B.~T., {Sergeyev}, A.~V., {et~al.} 2025, \aap, 700, A183, \dodoi{10.1051/0004-6361/202555633}

\bibitem[{{Beniyama} {et~al.}(2024){Beniyama}, {Sergeyev}, {Tholen}, \& {Micheli}}]{Beniyama2024}
{Beniyama}, J., {Sergeyev}, A.~V., {Tholen}, D.~J., \& {Micheli}, M. 2024, \aap, 690, A180, \dodoi{10.1051/0004-6361/202451414}

\bibitem[{{Beniyama} {et~al.}(2022){Beniyama}, {Sako}, {Ohsawa}, {Takita}, {Kobayashi}, {Okumura}, {Urakawa}, {Yoshikawa}, {Usui}, {Yoshida}, {Doi}, {Niino}, {Shigeyama}, {Tanaka}, {Tominaga}, {Aoki}, {Arima}, {Arimatsu}, {Kasuga}, {Kondo}, {Mori}, {Takahashi}, \& {Watanabe}}]{Beniyama2022}
{Beniyama}, J., {Sako}, S., {Ohsawa}, R., {et~al.} 2022, \pasj, 74, 877, \dodoi{10.1093/pasj/psac043}

\bibitem[{{Beniyama} {et~al.}(2023{\natexlab{a}}){Beniyama}, {Sekiguchi}, {Kuroda}, {Arai}, {Ishibashi}, {Ishiguro}, {Yoshida}, {Senshu}, {Ootsubo}, {Sako}, {Ohsawa}, {Takita}, {Geem}, \& {Bach}}]{Beniyama2023a}
{Beniyama}, J., {Sekiguchi}, T., {Kuroda}, D., {et~al.} 2023{\natexlab{a}}, \pasj, 75, 297, \dodoi{10.1093/pasj/psac109}

\bibitem[{{Beniyama} {et~al.}(2023{\natexlab{b}}){Beniyama}, {Sako}, {Ohtsuka}, {Sekiguchi}, {Ishiguro}, {Kuroda}, {Urakawa}, {Yoshida}, {Takumi}, {Maeda}, {Takahashi}, {Takagi}, {Saito}, {Nakaoka}, {Saito}, {Ohshima}, {Imazawa}, {Kagitani}, \& {Takita}}]{Beniyama2023b}
{Beniyama}, J., {Sako}, S., {Ohtsuka}, K., {et~al.} 2023{\natexlab{b}}, \apj, 955, 143, \dodoi{10.3847/1538-4357/ace88f}

\bibitem[{{Beniyama} {et~al.}(2023{\natexlab{c}}){Beniyama}, {Ohsawa}, {Avdellidou}, {Sako}, {Takita}, {Ishiguro}, {Sekiguchi}, {Usui}, {Kinoshita}, {Lee}, {Takumi}, {Ferrais}, \& {Jehin}}]{Beniyama2023c}
{Beniyama}, J., {Ohsawa}, R., {Avdellidou}, C., {et~al.} 2023{\natexlab{c}}, \aj, 166, 229, \dodoi{10.3847/1538-3881/ad0151}

\bibitem[{{Binzel} {et~al.}(2004){Binzel}, {Perozzi}, {Rivkin}, {Rossi}, {Harris}, {Bus}, {Valsecchi}, \& {Slivan}}]{Binzel2004}
{Binzel}, R.~P., {Perozzi}, E., {Rivkin}, A.~S., {et~al.} 2004, \maps, 39, 351, \dodoi{10.1111/j.1945-5100.2004.tb00098.x}

\bibitem[{Binzel {et~al.}(2019)Binzel, DeMeo, Turtelboom, Bus, Tokunaga, Burbine, Lantz, Polishook, Carry, Morbidelli, Birlan, Vernazza, Burt, Moskovitz, Slivan, Thomas, Rivkin, Hicks, Dunn, Reddy, Sanchez, Granvik, \& Kohout}]{Binzel2019}
Binzel, R.~P., DeMeo, F.~E., Turtelboom, E.~V., {et~al.} 2019, Icarus, 324, 41, \dodoi{10.1016/j.icarus.2018.12.035}

\bibitem[{{Birtwhistle}(2009)}]{Birtwhistle2009}
{Birtwhistle}, P. 2009, Minor Planet Bulletin, 36, 186

\bibitem[{{Birtwhistle}(2018{\natexlab{a}})}]{Birtwhistle2018a}
---. 2018{\natexlab{a}}, Minor Planet Bulletin, 45, 178

\bibitem[{{Birtwhistle}(2018{\natexlab{b}})}]{Birtwhistle2018b}
---. 2018{\natexlab{b}}, Minor Planet Bulletin, 45, 215

\bibitem[{{Birtwhistle}(2021{\natexlab{a}})}]{Birtwhistle2021a}
---. 2021{\natexlab{a}}, Minor Planet Bulletin, 48, 26

\bibitem[{{Birtwhistle}(2021{\natexlab{b}})}]{Birtwhistle2021b}
---. 2021{\natexlab{b}}, Minor Planet Bulletin, 48, 180

\bibitem[{{Birtwhistle}(2021{\natexlab{c}})}]{Birtwhistle2021c}
---. 2021{\natexlab{c}}, Minor Planet Bulletin, 48, 286

\bibitem[{{Birtwhistle}(2021{\natexlab{d}})}]{Birtwhistle2021d}
---. 2021{\natexlab{d}}, Minor Planet Bulletin, 48, 341

\bibitem[{{Birtwhistle}(2021{\natexlab{e}})}]{Birtwhistle2021e}
---. 2021{\natexlab{e}}, Minor Planet Bulletin, 48, 346

\bibitem[{{Birtwhistle}(2022{\natexlab{a}})}]{Birtwhistle2022a}
---. 2022{\natexlab{a}}, Minor Planet Bulletin, 49, 90

\bibitem[{{Birtwhistle}(2022{\natexlab{b}})}]{Birtwhistle2022b}
---. 2022{\natexlab{b}}, Minor Planet Bulletin, 49, 169

\bibitem[{{Birtwhistle}(2022{\natexlab{c}})}]{Birtwhistle2022c}
---. 2022{\natexlab{c}}, Minor Planet Bulletin, 49, 280

\bibitem[{{Birtwhistle}(2023{\natexlab{a}})}]{Birtwhistle2023a}
---. 2023{\natexlab{a}}, Minor Planet Bulletin, 50, 26

\bibitem[{{Birtwhistle}(2023{\natexlab{b}})}]{Birtwhistle2023b}
---. 2023{\natexlab{b}}, Minor Planet Bulletin, 50, 131

\bibitem[{{Birtwhistle}(2023{\natexlab{c}})}]{Birtwhistle2023c}
---. 2023{\natexlab{c}}, Minor Planet Bulletin, 50, 202

\bibitem[{{Birtwhistle}(2023{\natexlab{d}})}]{Birtwhistle2023d}
---. 2023{\natexlab{d}}, Minor Planet Bulletin, 50, 295

\bibitem[{{Birtwhistle}(2024{\natexlab{a}})}]{Birtwhistle2024a}
---. 2024{\natexlab{a}}, Minor Planet Bulletin, 51, 66

\bibitem[{{Birtwhistle}(2024{\natexlab{b}})}]{Birtwhistle2024b}
---. 2024{\natexlab{b}}, Minor Planet Bulletin, 51, 184

\bibitem[{{Birtwhistle}(2024{\natexlab{c}})}]{Birtwhistle2024c}
---. 2024{\natexlab{c}}, Minor Planet Bulletin, 51, 277

\bibitem[{{Birtwhistle}(2024{\natexlab{d}})}]{Birtwhistle2024d}
---. 2024{\natexlab{d}}, Minor Planet Bulletin, 51, 363

\bibitem[{{Birtwhistle}(2025{\natexlab{a}})}]{Birtwhistle2025a}
---. 2025{\natexlab{a}}, Minor Planet Bulletin, 52, 59

\bibitem[{{Birtwhistle}(2025{\natexlab{b}})}]{Birtwhistle2025b}
---. 2025{\natexlab{b}}, Minor Planet Bulletin, 52, 153

\bibitem[{{Birtwhistle}(2025{\natexlab{c}})}]{Birtwhistle2025c}
---. 2025{\natexlab{c}}, Minor Planet Bulletin, 52, 256

\bibitem[{{Birtwhistle} \& {Masiero}(2011{\natexlab{a}})}]{Birtwhistle2011a}
{Birtwhistle}, P., \& {Masiero}, J. 2011{\natexlab{a}}, Minor Planet Bulletin, 38, 36

\bibitem[{{Birtwhistle} \& {Masiero}(2011{\natexlab{b}})}]{Birtwhistle2011b}
---. 2011{\natexlab{b}}, Minor Planet Bulletin, 38, 36

\bibitem[{{Bolin} {et~al.}(2024){Bolin}, {Ghosal}, \& {Jedicke}}]{Bolin2024}
{Bolin}, B.~T., {Ghosal}, M., \& {Jedicke}, R. 2024, \mnras, 527, 1633, \dodoi{10.1093/mnras/stad3227}

\bibitem[{{Bowell} {et~al.}(1989){Bowell}, {Hapke}, {Domingue}, {Lumme}, {Peltoniemi}, \& {Harris}}]{Bowell1989}
{Bowell}, E., {Hapke}, B., {Domingue}, D., {et~al.} 1989, in Asteroids II, ed. R.~P. {Binzel}, T.~{Gehrels}, \& M.~S. {Matthews} (Tucson, AZ: Univ. Arizona Press), 524--556

\bibitem[{{Carbognani}(2017)}]{Carbognani2017}
{Carbognani}, A. 2017, \planss, 147, 1, \dodoi{10.1016/j.pss.2017.07.019}

\bibitem[{{Chambers} {et~al.}(2016){Chambers}, {Magnier}, {Metcalfe}, {Flewelling}, {Huber}, {Waters}, {Denneau}, {Draper}, {Farrow}, {Finkbeiner}, {Holmberg}, {Koppenhoefer}, {Price}, {Rest}, {Saglia}, {Schlafly}, {Smartt}, {Sweeney}, {Wainscoat}, {Burgett}, {Chastel}, {Grav}, {Heasley}, {Hodapp}, {Jedicke}, {Kaiser}, {Kudritzki}, {Luppino}, {Lupton}, {Monet}, {Morgan}, {Onaka}, {Shiao}, {Stubbs}, {Tonry}, {White}, {Ba{\~n}ados}, {Bell}, {Bender}, {Bernard}, {Boegner}, {Boffi}, {Botticella}, {Calamida}, {Casertano}, {Chen}, {Chen}, {Cole}, {Deacon}, {Frenk}, {Fitzsimmons}, {Gezari}, {Gibbs}, {Goessl}, {Goggia}, {Gourgue}, {Goldman}, {Grant}, {Grebel}, {Hambly}, {Hasinger}, {Heavens}, {Heckman}, {Henderson}, {Henning}, {Holman}, {Hopp}, {Ip}, {Isani}, {Jackson}, {Keyes}, {Koekemoer}, {Kotak}, {Le}, {Liska}, {Long}, {Lucey}, {Liu}, {Martin}, {Masci}, {McLean}, {Mindel}, {Misra}, {Morganson}, {Murphy}, {Obaika}, {Narayan}, {Nieto-Santisteban}, {Norberg}, {Peacock}, {Pier}, {Postman}, {Primak}, {Rae}, {Rai},
  {Riess}, {Riffeser}, {Rix}, {R{\"o}ser}, {Russel}, {Rutz}, {Schilbach}, {Schultz}, {Scolnic}, {Strolger}, {Szalay}, {Seitz}, {Small}, {Smith}, {Soderblom}, {Taylor}, {Thomson}, {Taylor}, {Thakar}, {Thiel}, {Thilker}, {Unger}, {Urata}, {Valenti}, {Wagner}, {Walder}, {Walter}, {Watters}, {Werner}, {Wood-Vasey}, \& {Wyse}}]{Chambers2016}
{Chambers}, K.~C., {Magnier}, E.~A., {Metcalfe}, N., {et~al.} 2016, arXiv:1612.05560, arXiv:1612.05560.
\newblock \doarXiv{1612.05560}

\bibitem[{{Chang} {et~al.}(2016){Chang}, {Lin}, {Ip}, {Prince}, {Kulkarni}, {Levitan}, {Laher}, \& {Surace}}]{Chang2016}
{Chang}, C.-K., {Lin}, H.-W., {Ip}, W.-H., {et~al.} 2016, \apjs, 227, 20, \dodoi{10.3847/0067-0049/227/2/20}

\bibitem[{{Chang} {et~al.}(2014){Chang}, {Waszczak}, {Lin}, {Ip}, {Prince}, {Kulkarni}, {Laher}, \& {Surace}}]{Chang2014b}
{Chang}, C.-K., {Waszczak}, A., {Lin}, H.-W., {et~al.} 2014, \apjl, 791, L35, \dodoi{10.1088/2041-8205/791/2/L35}

\bibitem[{{Chang} {et~al.}(2017){Chang}, {Lin}, {Ip}, {Lin}, {Kupfer}, {Prince}, {Ye}, {Laher}, {Lee}, \& {Moon}}]{Chang2017}
{Chang}, C.-K., {Lin}, H.-W., {Ip}, W.-H., {et~al.} 2017, \apjl, 840, L22, \dodoi{10.3847/2041-8213/aa6ff5}

\bibitem[{{Chang} {et~al.}(2019){Chang}, {Lin}, {Ip}, {Chen}, {Yeh}, {Chambers}, {Magnier}, {Huber}, {Flewelling}, {Waters}, {Wainscoat}, \& {Schultz}}]{Chang2019}
---. 2019, \apjs, 241, 6, \dodoi{10.3847/1538-4365/ab01fe}

\bibitem[{{Chang} {et~al.}(2021){Chang}, {Chen}, {Fraser}, {Yoshida}, {Lehner}, {Wang}, {Kavelaars}, {Pike}, {Alexandersen}, {Ito}, {Choi}, {Granados Contreras}, {Jeongahn}, {Ji}, {Kim}, {Lawler}, {Li}, {Lin}, {Sofia Lykawka}, {Moon}, {More}, {Mu{\~n}oz-Guti{\'e}rrez}, {Ohtsuki}, {Terai}, {Urakawa}, {Zhang}, {Zhao}, {Zhou}, \& {Fossil Collaboration}}]{Chang2021}
{Chang}, C.-K., {Chen}, Y.-T., {Fraser}, W.~C., {et~al.} 2021, \psj, 2, 191, \dodoi{10.3847/PSJ/ac13a4}

\bibitem[{{Chang} {et~al.}(2022{\natexlab{a}}){Chang}, {Yeh}, {Tan}, {Ip}, {Kelley}, {Ye}, {Lin}, {Ngeow}, {Bolin}, {Prince}, {Bellm}, {Dekany}, {Duev}, {Graham}, \& {Zwicky Transient Facility Collaboration}}]{Chang2022b}
{Chang}, C.-K., {Yeh}, T.-S., {Tan}, H., {et~al.} 2022{\natexlab{a}}, \apjl, 932, L5, \dodoi{10.3847/2041-8213/ac6e5e}

\bibitem[{{Chang} {et~al.}(2022{\natexlab{b}}){Chang}, {Chen}, {Fraser}, {Lehner}, {Wang}, {Alexandersen}, {Choi}, {Granados Contreras}, {Ito}, {Jeongahn}, {Ji}, {Kavelaars}, {Kim}, {Lawler}, {Li}, {Lin}, {Lykawka}, {Moon}, {More}, {Mu{\~n}oz-Guti{\'e}rrez}, {Ohtsuki}, {Pike}, {Terai}, {Urakawa}, {Yoshida}, {Zhang}, {Zhao}, {Zhou}, \& {(The Fossil Collaboration)}}]{Chang2022a}
{Chang}, C.-K., {Chen}, Y.-T., {Fraser}, W.~C., {et~al.} 2022{\natexlab{b}}, \apjs, 259, 7, \dodoi{10.3847/1538-4365/ac50ac}

\bibitem[{{Degewij} {et~al.}(1979){Degewij}, {Tedesco}, \& {Zellner}}]{Degewij1979}
{Degewij}, J., {Tedesco}, E.~F., \& {Zellner}, B. 1979, \icarus, 40, 364, \dodoi{10.1016/0019-1035(79)90029-0}

\bibitem[{{Devog{\`e}le} {et~al.}(2019){Devog{\`e}le}, {Moskovitz}, {Thirouin}, {Gustaffson}, {Magnuson}, {Thomas}, {Willman}, {Christensen}, {Person}, {Binzel}, {Polishook}, {DeMeo}, {Hinkle}, {Trilling}, {Mommert}, {Burt}, \& {Skiff}}]{Devogele2019}
{Devog{\`e}le}, M., {Moskovitz}, N., {Thirouin}, A., {et~al.} 2019, \aj, 158, 196, \dodoi{10.3847/1538-3881/ab43dd}

\bibitem[{{Devog{\`e}le} {et~al.}(2024{\natexlab{a}}){Devog{\`e}le}, {Buzzi}, {Micheli}, {Cano}, {Conversi}, {Jehin}, {Ferrais}, {Oca{\~n}a}, {F{\"o}hring}, {Drury}, {Benkhaldoun}, \& {Jenniskens}}]{Devogele2024b}
{Devog{\`e}le}, M., {Buzzi}, L., {Micheli}, M., {et~al.} 2024{\natexlab{a}}, \aap, 689, A211, \dodoi{10.1051/0004-6361/202450263}

\bibitem[{{Devog{\`e}le} {et~al.}(2024{\natexlab{b}}){Devog{\`e}le}, {McGilvray}, {MacLennan}, {Monchinski}, {Marshall}, {Hickson}, {Virkki}, {Giorgini}, {Abe}, {Augustin}, {Aznar-Mac{\'\i}as}, {Baudouin}, {Behrend}, {Bendjoya}, {Benkhaldoun}, {Bosch}, {Cellino}, {Chatelain}, {Deldem}, {Ferrais}, {Goncalves}, {Houdin}, {Hus{\'a}rik}, {Jehin}, {Kareta}, {Kim}, {Licandro}, {Lister}, {Medeiros}, {Pravec}, {Rivet}, {Rousseau}, {Roh}, {Skiff}, {Taylor}, {Venditti}, {Vernet}, {Vienney}, {Yim}, \& {Zambrano-Marin}}]{Devogele2024a}
{Devog{\`e}le}, M., {McGilvray}, A., {MacLennan}, E., {et~al.} 2024{\natexlab{b}}, \psj, 5, 44, \dodoi{10.3847/PSJ/ad1f70}

\bibitem[{{Ginsburg} {et~al.}(2019){Ginsburg}, {Sip{\H{o}}cz}, {Brasseur}, {Cowperthwaite}, {Craig}, {Deil}, {Guillochon}, {Guzman}, {Liedtke}, {Lian Lim}, {Lockhart}, {Mommert}, {Morris}, {Norman}, {Parikh}, {Persson}, {Robitaille}, {Segovia}, {Singer}, {Tollerud}, {de Val-Borro}, {Valtchanov}, {Woillez}, {Astroquery Collaboration}, \& {a subset of astropy Collaboration}}]{Ginsburg2019}
{Ginsburg}, A., {Sip{\H{o}}cz}, B.~M., {Brasseur}, C.~E., {et~al.} 2019, \aj, 157, 98, \dodoi{10.3847/1538-3881/aafc33}

\bibitem[{{Greenstreet} {et~al.}(2026){Greenstreet}, {Li}, {Vavilov}, {Singh}, {Juri{\'c}}, {Ivezi{\'c}}, {Eggl}, {Koumjian}, {Moeyens}, {Carruba}, {Womack}, {Granvik}, {Alexov}, {Antilogus}, {Bauman{\'c}}, {Bellm}, {Boucaud}, {Bradshaw}, {Carlin}, {Chiang}, {Daly}, {Daruich}, {Daubard}, {Dennihy}, {Deppe}, {Drass}, {Gangler}, {Le Guillou}, {Guy}, {Hascall}, {Ingraham}, {Jee}, {Jenness}, {Kahn}, {Kannawadi}, {Kelvin}, {Kurlander}, {Laporte}, {Lust}, {Lutfi}, {MacArthur}, {Mainetti}, {Marc}, {Plazas Malag{\'o}n}, {Jim{\'e}nez Mej{\'\i}as}, {Menanteau}, {Mills}, {O'Mullane}, {Neto}, {Neveu}, {Nourbakhsh}, {Park}, {Patterson}, {Peterson}, {Quint}, {Ribeiro}, {Ridgway}, {van Reeven}, {Sebag}, {Sedaghat}, {Shaw}, {Strauss}, {Suberlak}, {Sullivan}, {Swinbank}, {Thomas}, {Thornton}, {Wood-Vasey}, {Walter}, {Ward}, \& {Willman}}]{Greenstreet2026}
{Greenstreet}, S., {Li}, Z.~C., {Vavilov}, D.~E., {et~al.} 2026, \apjl, 996, L33, \dodoi{10.3847/2041-8213/ae2a30}

\bibitem[{{Guti{\'e}rrez} {et~al.}(2006){Guti{\'e}rrez}, {Davidsson}, {Ortiz}, {Rodrigo}, \& {Vidal-Nu{\~n}ez}}]{Gutierrez2006}
{Guti{\'e}rrez}, P.~J., {Davidsson}, B.~J.~R., {Ortiz}, J.~L., {Rodrigo}, R., \& {Vidal-Nu{\~n}ez}, M.~J. 2006, \aap, 454, 367, \dodoi{10.1051/0004-6361:20064838}

\bibitem[{{Harris} {et~al.}(2020){Harris}, {Millman}, {van der Walt}, {Gommers}, {Virtanen}, {Cournapeau}, {Wieser}, {Taylor}, {Berg}, {Smith}, {Kern}, {Picus}, {Hoyer}, {van Kerkwijk}, {Brett}, {Haldane}, {del R{\'\i}o}, {Wiebe}, {Peterson}, {G{\'e}rard-Marchant}, {Sheppard}, {Reddy}, {Weckesser}, {Abbasi}, {Gohlke}, \& {Oliphant}}]{Harris2020}
{Harris}, C.~R., {Millman}, K.~J., {van der Walt}, S.~J., {et~al.} 2020, \nat, 585, 357, \dodoi{10.1038/s41586-020-2649-2}

\bibitem[{{Hasegawa} {et~al.}(2024){Hasegawa}, {Marsset}, {DeMeo}, {Hanu{\v{s}}}, {Binzel}, {Bus}, {Burt}, {Polishook}, {Thomas}, {Geem}, {Ishiguro}, {Kuroda}, \& {Vernazza}}]{Hasegawa2024}
{Hasegawa}, S., {Marsset}, M., {DeMeo}, F.~E., {et~al.} 2024, \aj, 167, 224, \dodoi{10.3847/1538-3881/ad3045}

\bibitem[{{Hatch} \& {Wiegert}(2015)}]{Hatch2015}
{Hatch}, P., \& {Wiegert}, P.~A. 2015, \planss, 111, 100, \dodoi{10.1016/j.pss.2015.03.019}

\bibitem[{{Hirabayashi} {et~al.}(2021){Hirabayashi}, {Mimasu}, {Sakatani}, {Watanabe}, {Tsuda}, {Saiki}, {Kikuchi}, {Kouyama}, {Yoshikawa}, {Tanaka}, {Nakazawa}, {Takei}, {Terui}, {Takeuchi}, {Fujii}, {Iwata}, {Tsumura}, {Matsuura}, {Shimaki}, {Urakawa}, {Ishibashi}, {Hasegawa}, {Ishiguro}, {Kuroda}, {Okumura}, {Sugita}, {Okada}, {Kameda}, {Kamata}, {Higuchi}, {Senshu}, {Noda}, {Matsumoto}, {Suetsugu}, {Hirai}, {Kitazato}, {Farnocchia}, {Naidu}, {Tholen}, {Hergenrother}, {Whiteley}, {Moskovitz}, {Abell}, \& {the Hayabusa2 extended mission study group}}]{Hirabayashi2021}
{Hirabayashi}, M., {Mimasu}, Y., {Sakatani}, N., {et~al.} 2021, Advances in Space Research, 68, 1533, \dodoi{10.1016/j.asr.2021.03.030}

\bibitem[{{Holsapple}(2004)}]{Holsapple2004}
{Holsapple}, K.~A. 2004, \icarus, 172, 272, \dodoi{10.1016/j.icarus.2004.05.023}

\bibitem[{Holsapple(2007)}]{Holsapple2007}
Holsapple, K.~A. 2007, Icarus, 187, 500, \dodoi{10.1016/j.icarus.2006.08.012}

\bibitem[{{Jenniskens} {et~al.}(2009){Jenniskens}, {Shaddad}, {Numan}, {Elsir}, {Kudoda}, {Zolensky}, {Le}, {Robinson}, {Friedrich}, {Rumble}, {Steele}, {Chesley}, {Fitzsimmons}, {Duddy}, {Hsieh}, {Ramsay}, {Brown}, {Edwards}, {Tagliaferri}, {Boslough}, {Spalding}, {Dantowitz}, {Kozubal}, {Pravec}, {Borovicka}, {Charvat}, {Vaubaillon}, {Kuiper}, {Albers}, {Bishop}, {Mancinelli}, {Sandford}, {Milam}, {Nuevo}, \& {Worden}}]{Jenniskens2009}
{Jenniskens}, P., {Shaddad}, M.~H., {Numan}, D., {et~al.} 2009, \nat, 458, 485, \dodoi{10.1038/nature07920}

\bibitem[{{Jenniskens} {et~al.}(2010){Jenniskens}, {Vaubaillon}, {Binzel}, {DeMeo}, {Nesvorn{\'y}}, {Bottke}, {Fitzsimmons}, {Hiroi}, {Marchis}, {Bishop}, {Vernazza}, {Zolensky}, {Herrin}, {Welten}, {Meier}, \& {Shaddad}}]{Jenniskens2010}
{Jenniskens}, P., {Vaubaillon}, J., {Binzel}, R.~P., {et~al.} 2010, \maps, 45, 1590, \dodoi{10.1111/j.1945-5100.2010.01153.x}

\bibitem[{{Kareta} {et~al.}(2024){Kareta}, {Vida}, {Micheli}, {Moskovitz}, {Wiegert}, {Brown}, {McCausland}, {Devillepoix}, {Male{\v{c}}i{\'c}}, {Prtenjak}, {{\v{S}}egon}, {Shafransky}, \& {Farnocchia}}]{Kareta2024b}
{Kareta}, T., {Vida}, D., {Micheli}, M., {et~al.} 2024, \psj, 5, 253, \dodoi{10.3847/PSJ/ad8b22}

\bibitem[{{Kikuchi} {et~al.}(2023){Kikuchi}, {Mimasu}, {Takei}, {Saiki}, {Scheeres}, {Hirabayashi}, {Wada}, {Yoshikawa}, {Watanabe}, {Tanaka}, \& {Tsuda}}]{Kikuchi2023}
{Kikuchi}, S., {Mimasu}, Y., {Takei}, Y., {et~al.} 2023, Acta Astronautica, 211, 295, \dodoi{10.1016/j.actaastro.2023.06.010}

\bibitem[{{Kiss} {et~al.}(2025){Kiss}, {Tak{\'a}cs}, {Kalup}, {Szak{\'a}ts}, {Moln{\'a}r}, {Plachy}, {S{\'a}rneczky}, {Szab{\'o}}, {Szab{\'o}}, {B{\'o}di}, \& {P{\'a}l}}]{Kiss2025}
{Kiss}, C., {Tak{\'a}cs}, N., {Kalup}, C.~E., {et~al.} 2025, \aap, 694, L17, \dodoi{10.1051/0004-6361/202453509}

\bibitem[{{Kurita} {et~al.}(2020){Kurita}, {Kino}, {Iwamuro}, {Ohta}, {Nogami}, {Izumiura}, {Yoshida}, {Matsubayashi}, {Kuroda}, {Nakatani}, {Yamamoto}, {Tsutsui}, {Iribe}, {Jikuya}, {Ohtani}, {Shibata}, {Takahashi}, {Tokoro}, {Maihara}, \& {Nagata}}]{Kurita2020}
{Kurita}, M., {Kino}, M., {Iwamuro}, F., {et~al.} 2020, \pasj, 72, 48, \dodoi{10.1093/pasj/psaa036}

\bibitem[{{Kwiatkowski} {et~al.}(2010){Kwiatkowski}, {Polinska}, {Loaring}, {Buckley}, {O'Donoghue}, {Kniazev}, \& {Romero Colmenero}}]{Kwiatkowski2010c}
{Kwiatkowski}, T., {Polinska}, M., {Loaring}, N., {et~al.} 2010, \aap, 511, A49, \dodoi{10.1051/0004-6361/200913468}

\bibitem[{{Lacerda} {et~al.}(2008){Lacerda}, {Jewitt}, \& {Peixinho}}]{Lacerda2008}
{Lacerda}, P., {Jewitt}, D., \& {Peixinho}, N. 2008, \aj, 135, 1749, \dodoi{10.1088/0004-6256/135/5/1749}

\bibitem[{{Lang} {et~al.}(2010){Lang}, {Hogg}, {Mierle}, {Blanton}, \& {Roweis}}]{Lang2010}
{Lang}, D., {Hogg}, D.~W., {Mierle}, K., {Blanton}, M., \& {Roweis}, S. 2010, \aj, 139, 1782, \dodoi{10.1088/0004-6256/139/5/1782}

\bibitem[{{Licandro} {et~al.}(2023){Licandro}, {Popescu}, {Tatsumi}, {Alarcon}, {Serra-Ricart}, {Medeiros}, {Morate}, {Tinaut-Ruano}, \& {de Le{\'o}n}}]{Licandro2023}
{Licandro}, J., {Popescu}, M., {Tatsumi}, E., {et~al.} 2023, \mnras, 521, 3784, \dodoi{10.1093/mnras/stad708}

\bibitem[{{Lomb}(1976)}]{Lomb1976}
{Lomb}, N.~R. 1976, \apss, 39, 447, \dodoi{10.1007/BF00648343}

\bibitem[{{L{\'o}pez-Oquendo} {et~al.}(2022){L{\'o}pez-Oquendo}, {Trilling}, {Gustafsson}, {Virkki}, {Rivera-Valent{\'\i}n}, {Granvik}, {Chandler}, {Chatelain}, {Taylor}, \& {Fernanda-Zambrano}}]{Lopez-Oquendo2022}
{L{\'o}pez-Oquendo}, A., {Trilling}, D.~E., {Gustafsson}, A., {et~al.} 2022, \psj, 3, 189, \dodoi{10.3847/PSJ/ac7e4f}

\bibitem[{{Marsset} {et~al.}(2022){Marsset}, {DeMeo}, {Burt}, {Polishook}, {Binzel}, {Granvik}, {Vernazza}, {Carry}, {Bus}, {Slivan}, {Thomas}, {Moskovitz}, \& {Rivkin}}]{Marsset2022a}
{Marsset}, M., {DeMeo}, F.~E., {Burt}, B., {et~al.} 2022, \aj, 163, 165, \dodoi{10.3847/1538-3881/ac532f}

\bibitem[{{McCully} {et~al.}(2018){McCully}, {Crawford}, {Kovacs}, {Tollerud}, {Betts}, {Bradley}, {Craig}, {Turner}, {Streicher}, {Sipocz}, {Robitaille}, \& {Deil}}]{McCully2018}
{McCully}, C., {Crawford}, S., {Kovacs}, G., {et~al.} 2018, {Astropy/Astroscrappy: V1.0.5 Zenodo Release}, v1.0.5, Zenodo,  Zenodo, \dodoi{10.5281/zenodo.1482019}

\bibitem[{Mommert {et~al.}(2016)Mommert, Trilling, Borth, Jedicke, Butler, Reyes-Ruiz, Pichardo, Petersen, Axelrod, \& Moskovitz}]{Mommert2016}
Mommert, M., Trilling, D.~E., Borth, D., {et~al.} 2016, The Astronomical Journal, 151, 98, \dodoi{10.3847/0004-6256/151/4/98}

\bibitem[{{Moskovitz} {et~al.}(2026){Moskovitz}, {Kareta}, {Hemmelgarn}, {Zigo}, {Devog{\`e}le}, {Thirouin}, {Breeland-Newcomb}, {Burt}, {Gustaffson}, {Magnuson}, {Mommert}, {Polishook}, {Schottland}, {Skiff}, {Thomas}, \& {Willman}}]{Moskovitz2026}
{Moskovitz}, N., {Kareta}, T., {Hemmelgarn}, S., {et~al.} 2026, arXiv e-prints, arXiv:2603.15872.
\newblock \doarXiv{2603.15872}

\bibitem[{{Novakovi{\'c}} \& {Guti{\'e}rrez}(2025)}]{Novakovic2025}
{Novakovi{\'c}}, B., \& {Guti{\'e}rrez}, P.~J. 2025, \aj, 170, 248, \dodoi{10.3847/1538-3881/ae0473}

\bibitem[{Oliphant(2015)}]{Oliphant2015}
Oliphant, T.~E. 2015, Guide to NumPy, 2nd edn. (North Charleston, SC, USA: CreateSpace Independent Publishing Platform)

\bibitem[{{Perna} {et~al.}(2016){Perna}, {Dotto}, {Ieva}, {Barucci}, {Bernardi}, {Fornasier}, {De Luise}, {Perozzi}, {Rossi}, {Mazzotta Epifani}, {Micheli}, \& {Deshapriya}}]{Perna2016}
{Perna}, D., {Dotto}, E., {Ieva}, S., {et~al.} 2016, \aj, 151, 11, \dodoi{10.3847/0004-6256/151/1/11}

\bibitem[{{Polishook} {et~al.}(2012){Polishook}, {Ofek}, {Waszczak}, {Kulkarni}, {Gal-Yam}, {Aharonson}, {Laher}, {Surace}, {Klein}, {Bloom}, {Brosch}, {Prialnik}, {Grillmair}, {Cenko}, {Kasliwal}, {Law}, {Levitan}, {Nugent}, {Poznanski}, \& {Quimby}}]{Polishook2012}
{Polishook}, D., {Ofek}, E.~O., {Waszczak}, A., {et~al.} 2012, \mnras, 421, 2094, \dodoi{10.1111/j.1365-2966.2012.20462.x}

\bibitem[{{Pravec} \& {Harris}(2000)}]{Pravec2000b}
{Pravec}, P., \& {Harris}, A.~W. 2000, \icarus, 148, 12, \dodoi{10.1006/icar.2000.6482}

\bibitem[{{Reback} {et~al.}(2021){Reback}, {jbrockmendel}, {McKinney}, {Van den Bossche}, {Augspurger}, {Cloud}, {Hawkins}, {gfyoung}, {Roeschke}, {Sinhrks}, {Klein}, {Petersen}, {Tratner}, {She}, {Ayd}, {Hoefler}, {Naveh}, {Garcia}, {Schendel}, {Hayden}, {Saxton}, {Darbyshire}, {Shadrach}, {Gorelli}, {Li}, {Zeitlin}, {Jancauskas}, {McMaster}, {Battiston}, \& {Seabold}}]{Reback2021}
{Reback}, J., {jbrockmendel}, {McKinney}, W., {et~al.} 2021, {pandas-dev/pandas: Pandas 1.3.4}, v1.3.4, Zenodo,  Zenodo, \dodoi{10.5281/zenodo.5574486}

\bibitem[{{Rond{\'o}n} {et~al.}(2020){Rond{\'o}n}, {Lazzaro}, {Rodrigues}, {Carvano}, {Roig}, {Monteiro}, {Arcoverde}, {Medeiros}, {Silva}, {Jasmim}, {Pr{\'a}}, {Hasselmann}, {Ribeiro}, {D{\'a}valos}, \& {Souza}}]{Rondon2020}
{Rond{\'o}n}, E., {Lazzaro}, D., {Rodrigues}, T., {et~al.} 2020, \pasp, 132, 065001, \dodoi{10.1088/1538-3873/ab87a7}

\bibitem[{{Sanchez} {et~al.}(2024){Sanchez}, {Reddy}, {Thirouin}, {Bottke}, {Kareta}, {De Florio}, {Sharkey}, {Battle}, {Cantillo}, \& {Pearson}}]{Sanchez2024}
{Sanchez}, J.~A., {Reddy}, V., {Thirouin}, A., {et~al.} 2024, \psj, 5, 131, \dodoi{10.3847/PSJ/ad445f}

\bibitem[{{S{\'a}nchez} \& {Scheeres}(2014)}]{Sanchez2014}
{S{\'a}nchez}, P., \& {Scheeres}, D.~J. 2014, \maps, 49, 788, \dodoi{10.1111/maps.12293}

\bibitem[{{Scargle}(1982)}]{Scargle1982}
{Scargle}, J.~D. 1982, \apj, 263, 835, \dodoi{10.1086/160554}

\bibitem[{{Sergeyev} \& {Carry}(2021)}]{Sergeyev2021}
{Sergeyev}, A.~V., \& {Carry}, B. 2021, \aap, 652, A59, \dodoi{10.1051/0004-6361/202140430}

\bibitem[{{Shepard} {et~al.}(2008){Shepard}, {Clark}, {Nolan}, {Benner}, {Ostro}, {Giorgini}, {Vilas}, {Jarvis}, {Lederer}, {Lim}, {McConnochie}, {Bell}, {Margot}, {Rivkin}, {Magri}, {Scheeres}, \& {Pravec}}]{Shepard2008}
{Shepard}, M.~K., {Clark}, B.~E., {Nolan}, M.~C., {et~al.} 2008, \icarus, 193, 20, \dodoi{10.1016/j.icarus.2007.09.006}

\bibitem[{{Strauss} {et~al.}(2024){Strauss}, {McNeill}, {Trilling}, {Valdes}, {Bernardinelli}, {Fuentes}, {Gerdes}, {Holman}, {Juri{\'c}}, {Lin}, {Markwardt}, {Mommert}, {Napier}, {Oldroyd}, {Payne}, {Rivkin}, {Schlichting}, {Sheppard}, {Smotherman}, {Trujillo}, {Adams}, \& {Chandler}}]{Strauss2024}
{Strauss}, R., {McNeill}, A., {Trilling}, D.~E., {et~al.} 2024, \aj, 168, 184, \dodoi{10.3847/1538-3881/ad7366}

\bibitem[{{Szab{\'o}} {et~al.}(2004){Szab{\'o}}, {Ivezi{\'c}}, {Juri{\'c}}, {Lupton}, \& {Kiss}}]{Szabo2004}
{Szab{\'o}}, G.~M., {Ivezi{\'c}}, {\v{Z}}., {Juri{\'c}}, M., {Lupton}, R., \& {Kiss}, L.~L. 2004, \mnras, 348, 987, \dodoi{10.1111/j.1365-2966.2004.07426.x}

\bibitem[{{Tak{\'a}cs} {et~al.}(2025){Tak{\'a}cs}, {Kiss}, {Szak{\'a}ts}, {Plachy}, {Kalup}, {Szab{\'o}}, {Moln{\'a}r}, {S{\'a}rneczky}, {Szab{\'o}}, {B{\'o}di}, \& {P{\'a}l}}]{Takacs2025}
{Tak{\'a}cs}, N., {Kiss}, C., {Szak{\'a}ts}, R., {et~al.} 2025, \apjl, 986, L33, \dodoi{10.3847/2041-8213/ade05b}

\bibitem[{{Tatsumi} {et~al.}(2020){Tatsumi}, {Domingue}, {Schr{\"o}der}, {Yokota}, {Kuroda}, {Ishiguro}, {Hasegawa}, {Hiroi}, {Honda}, {Hemmi}, {Le Corre}, {Sakatani}, {Morota}, {Yamada}, {Kameda}, {Koyama}, {Suzuki}, {Cho}, {Yoshioka}, {Matsuoka}, {Honda}, {Hayakawa}, {Hirata}, {Hirata}, {Yamamoto}, {Vilas}, {Takato}, {Yoshikawa}, {Abe}, \& {Sugita}}]{Tatsumi2020}
{Tatsumi}, E., {Domingue}, D., {Schr{\"o}der}, S., {et~al.} 2020, \aap, 639, A83, \dodoi{10.1051/0004-6361/201937096}

\bibitem[{{Thirouin} {et~al.}(2016){Thirouin}, {Moskovitz}, {Binzel}, {Christensen}, {DeMeo}, {Person}, {Polishook}, {Thomas}, {Trilling}, {Willman}, {Hinkle}, {Burt}, {Avner}, \& {Aceituno}}]{Thirouin2016}
{Thirouin}, A., {Moskovitz}, N., {Binzel}, R.~P., {et~al.} 2016, \aj, 152, 163, \dodoi{10.3847/0004-6256/152/6/163}

\bibitem[{{Thirouin} {et~al.}(2018){Thirouin}, {Moskovitz}, {Binzel}, {Christensen}, {DeMeo}, {Person}, {Polishook}, {Thomas}, {Trilling}, {Willman}, {Burt}, {Hinkle}, \& {Pugh}}]{Thirouin2018}
{Thirouin}, A., {Moskovitz}, N.~A., {Binzel}, R.~P., {et~al.} 2018, \apjs, 239, 4, \dodoi{10.3847/1538-4365/aae1b0}

\bibitem[{{Tonry} {et~al.}(2012){Tonry}, {Stubbs}, {Lykke}, {Doherty}, {Shivvers}, {Burgett}, {Chambers}, {Hodapp}, {Kaiser}, {Kudritzki}, {Magnier}, {Morgan}, {Price}, \& {Wainscoat}}]{Tonry2012}
{Tonry}, J.~L., {Stubbs}, C.~W., {Lykke}, K.~R., {et~al.} 2012, \apj, 750, 99, \dodoi{10.1088/0004-637X/750/2/99}

\bibitem[{{van Dokkum}(2001)}]{vanDokkum2001}
{van Dokkum}, P.~G. 2001, \pasp, 113, 1420, \dodoi{10.1086/323894}

\bibitem[{{VanderPlas}(2018)}]{VanderPlas2018}
{VanderPlas}, J.~T. 2018, \apjs, 236, 16, \dodoi{10.3847/1538-4365/aab766}

\bibitem[{{Virtanen} {et~al.}(2020){Virtanen}, {Gommers}, {Oliphant}, {Haberland}, {Reddy}, {Cournapeau}, {Burovski}, {Peterson}, {Weckesser}, {Bright}, {van der Walt}, {Brett}, {Wilson}, {Millman}, {Mayorov}, {Nelson}, {Jones}, {Kern}, {Larson}, {Carey}, {Polat}, {Feng}, {Moore}, {VanderPlas}, {Laxalde}, {Perktold}, {Cimrman}, {Henriksen}, {Quintero}, {Harris}, {Archibald}, {Ribeiro}, {Pedregosa}, {van Mulbregt}, \& {SciPy 1. 0 Contributors}}]{Virtanen2020}
{Virtanen}, P., {Gommers}, R., {Oliphant}, T.~E., {et~al.} 2020, Nature Methods, 17, 261, \dodoi{10.1038/s41592-019-0686-2}

\bibitem[{Warner {et~al.}(2009)Warner, Harris, \& Pravec}]{Warner2009}
Warner, B.~D., Harris, A.~W., \& Pravec, P. 2009, Icarus, 202, 134, \dodoi{10.1016/j.icarus.2009.02.003}

\bibitem[{{Yeh} {et~al.}(2020){Yeh}, {Li}, {Chang}, {Zhao}, {Ji}, {Lin}, \& {Ip}}]{Yeh2020}
{Yeh}, T.-S., {Li}, B., {Chang}, C.-K., {et~al.} 2020, \aj, 160, 73, \dodoi{10.3847/1538-3881/ab9a32}

\bibitem[{{Zappala} {et~al.}(1990){Zappala}, {Cellino}, {Barucci}, {Fulchignoni}, \& {Lupishko}}]{Zappala1990}
{Zappala}, V., {Cellino}, A., {Barucci}, A.~M., {Fulchignoni}, M., \& {Lupishko}, D.~F. 1990, \aap, 231, 548

\bibitem[{{Zhang} {et~al.}(2021){Zhang}, {Xu}, \& {Ding}}]{Zhang2021}
{Zhang}, T., {Xu}, K., \& {Ding}, X. 2021, Nature Astronomy, 5, 730, \dodoi{10.1038/s41550-021-01418-9}

\end{thebibliography}
